\PassOptionsToPackage{unicode}{hyperref}
\PassOptionsToPackage{hyphens}{url}
\documentclass[
  11pt]{article}
\usepackage{amsmath,amssymb}
\usepackage{iftex}
\ifPDFTeX
  \usepackage[T1]{fontenc}
  \usepackage[utf8]{inputenc}
  \usepackage{textcomp} % provide euro and other symbols
\else % if luatex or xetex
  \usepackage{unicode-math} % this also loads fontspec
  \defaultfontfeatures{Scale=MatchLowercase}
  \defaultfontfeatures[\rmfamily]{Ligatures=TeX,Scale=1}
\fi
\IfFileExists{lmodern.sty}{\usepackage{lmodern}}{}
\ifPDFTeX\else
\fi
\IfFileExists{upquote.sty}{\usepackage{upquote}}{}
\IfFileExists{microtype.sty}{% use microtype if available
  \usepackage[expansion=false]{microtype}
  \UseMicrotypeSet[protrusion]{basicmath} % disable protrusion for tt fonts
}{}
\makeatletter
\@ifundefined{KOMAClassName}{% if non-KOMA class
  \IfFileExists{parskip.sty}{%
    \usepackage{parskip}
  }{% else
    \setlength{\parindent}{0pt}
    \setlength{\parskip}{6pt plus 2pt minus 1pt}}
}{% if KOMA class
  \KOMAoptions{parskip=half}}
\makeatother
\usepackage{xcolor}
\usepackage[margin=1in]{geometry}
\usepackage{graphicx}
\makeatletter
\def\maxwidth{\ifdim\Gin@nat@width>\linewidth\linewidth\else\Gin@nat@width\fi}
\def\maxheight{\ifdim\Gin@nat@height>\textheight\textheight\else\Gin@nat@height\fi}
\makeatother
\setkeys{Gin}{width=\maxwidth,height=\maxheight,keepaspectratio}
\makeatletter
\def\fps@figure{htbp}
\makeatother
\providecommand{\tightlist}{%
  \setlength{\itemsep}{0pt}\setlength{\parskip}{0pt}}
\newlength{\cslhangindent}
\newlength{\csllabelwidth}
\newlength{\cslentryspacingunit} % times entry-spacing
\newenvironment{CSLReferences}[2] % #1 hanging-ident, #2 entry spacing
 {% don't indent paragraphs
  \setlength{\parindent}{0pt}
  \ifodd #1
  \let\oldpar\par
  \def\par{\hangindent=\cslhangindent\oldpar}
  \fi
  \setlength{\parskip}{#2\cslentryspacingunit}
 }%
 {}
\usepackage{calc}

\ifLuaTeX
  \usepackage{selnolig}  % disable illegal ligatures
\fi
\IfFileExists{bookmark.sty}{\usepackage{bookmark}}{\usepackage{hyperref}}
\IfFileExists{xurl.sty}{\usepackage{xurl}}{} % add URL line breaks if available
\hypersetup{
  pdftitle={Reachability Does Not Imply Searchability in Expanding Networks},
  pdfauthor={Antonio Scala},
  hidelinks,
  pdfcreator={LaTeX via pandoc}}

\title{Reachability Does Not Imply Searchability in Expanding Networks}
\author{Antonio Scala}
\date{CNR-ISC, Rome, Italy}

\begin{document}
\maketitle

\textbf{Antonio Scala}\\
\emph{CNR-ISC, Rome, Italy}

\begin{center}\rule{0.5\linewidth}{0.5pt}\end{center}

\hypertarget{abstract}{%
\section{Abstract}\label{abstract}}

Routing and search respond in opposite ways to rapid network expansion.
Short graph distances make a known destination easy to reach, while the
neighborhood accessible within a few hops can be vastly larger than any
finite inspection budget. We show that this tension imposes an
algorithm-independent constraint on sparse search. If \(m_q\) relevant
nodes are placed without structural information among \(N\) nodes and at
most \(M\) nodes can be inspected, any target-blind exploration protocol
has success probability at most \(Mm_q/N\), irrespective of correlations
between successive inspections, of revisits, and of any structural
preference the exploration rule may have. Geometry determines when this
constraint becomes relevant: a target becomes accessible with order-one
probability when the graph-distance ball volume satisfies
\(V(R_c)\sim N/m_q\). At this scale, any inspection mechanism achieving
a fixed success probability \(\delta>0\) must enrich the probability of
inspecting relevant nodes by at least order \(N/(Mm_q)\) relative to the
neutral target density. For uniformly searched candidate sets, this
entails a vanishing visible fraction of the accessible region. Thus
rapidly expanding networks can make rare targets geometrically close
while target-blind search remains ineffective. Numerical results on
Krioukov hyperbolic random graphs and two-dimensional lattices show that
this separation can arise within only a few hops in the rapidly
expanding case, while the lattice accessibility radius grows
algebraically.

\begin{center}\rule{0.5\linewidth}{0.5pt}\end{center}

\hypertarget{introduction}{%
\section{1. Introduction}\label{introduction}}

Large complex networks are often described as navigable because distant
nodes can be reached along short paths. Hyperbolic network models
provide a geometric explanation: placing nodes in a negatively curved
latent space, together with heterogeneous connectivity, can generate
short distances and support efficient greedy routing (Krioukov et al.
2010; Boguñá, Krioukov, and Claffy 2009; Boguñá and Krioukov 2009). But
navigation and search answer different questions. Navigation asks how to
reach a known destination. Search asks how to find a relevant node when
its location is not known.

Kleinberg's algorithmic formulation of the small-world problem already
separated short paths from decentralized navigability. Short paths can
be abundant even when no decentralized algorithm using only local
information can find them efficiently; within the family of models he
studied, efficient decentralized routing emerges only for a specific
scaling of long-range links (Kleinberg 2000). Subsequent work showed how
structural cues can support Milgram-style routing in social-network
models (Lattanzi, Panconesi, and Sivakumar 2011). In both cases the
destination is specified in advance. Here we ask the complementary
question: what can finite-budget exploration achieve when the relevant
target is not known and the inspection rule has no information
correlated with relevance?

For sparse targets, the resulting constraint is elementary but
consequential. Suppose relevant nodes are rare. As the search radius
grows, the number of reachable nodes may increase very rapidly, so that
a relevant node soon lies nearby. Yet a finite search budget allows only
a small number of those reachable nodes to be inspected. Geometry
controls how many nodes are \emph{accessible}; attention controls how
many can actually be \emph{inspected}. The problem is therefore governed
by the competition between the growth of an accessible volume and a
finite sampling capacity.

The whole separation can be anticipated from two counts. If the fraction
of relevant nodes is \(m_q/N\), a ball must contain of order \(N/m_q\)
nodes before it typically contains a target. But a target-blind search
with only \(M\) inspection steps can examine at most a fraction

\begin{equation*}
\frac{M}{N/m_q}=\frac{Mm_q}{N}
\end{equation*}

of that characteristic accessible volume. The rest of this paper makes
the counting argument exact, establishes it for arbitrary correlated
exploration, and determines what a successful search must therefore
supply.

\hypertarget{accessibility}{%
\section{2. Accessibility}\label{accessibility}}

Let \(G=(\mathcal V,E)\) be a connected graph with \(N=|\mathcal V|\),
and fix a seed node \(x_0\). The graph-distance ball of radius \(R\) is

\begin{equation*}
B(x_0,R)=\{v\in\mathcal V:d(x_0,v)\le R\},
\qquad
V(R)=|B(x_0,R)|.
\end{equation*}

For a query \(q\), let \(T_q\subset\mathcal V\) denote the set of
relevant nodes, with \(|T_q|=m_q\). As a neutral baseline, the \(m_q\)
targets are placed uniformly among all subsets of \(\mathcal V\) of that
size, independently of graph structure. This is a null model rather than
a model of empirical relevance distributions: it deliberately removes
any relevance signal carried by graph structure. Section 4 shows that
the central bound survives a much wider class of correlated target
placements.

For a given target placement, let

\begin{equation*}
m_R=|T_q\cap B(x_0,R)|
\end{equation*}

be the number of targets inside the ball. Averaging over the neutral
target placement gives

\begin{equation}\label{eq:1}
\lambda_R
\equiv
\mathbb E[m_R]
=
\frac{m_qV(R)}{N}.
\end{equation}

The ball therefore contains of order one target on average when

\begin{equation}\label{eq:2}
V(R_c)\sim\frac{N}{m_q}.
\end{equation}

We call \(R_c\) the characteristic accessibility radius. The exact
probability that at least one target is inside the ball is

\begin{equation}\label{eq:3}
\Psi(R)
=
1-
\frac{\binom{N-V(R)}{m_q}}
     {\binom{N}{m_q}}.
\end{equation}

Because graph balls grow in discrete jumps, in the simulations we use
the first radius at which the mean target count reaches one,

\begin{equation}\label{eq:4}
R_c^{\rm op}
=
\min\left\{R:\frac{m_qV(R)}{N}\ge1\right\}.
\end{equation}

At this radius, \(\Psi(R)\) is already of order unity: the second-moment
bound in Appendix SI-3 gives \(\Psi\ge\lambda_R/(1+\lambda_R)\ge1/2\)
whenever \(\lambda_R\ge1\). Accessibility is therefore controlled by the
growth of the graph-distance volume \(V(R)\).

\hypertarget{the-target-blind-bound}{%
\section{3. The target-blind bound}\label{the-target-blind-bound}}

\hypertarget{distinct-coverage}{%
\subsection{3.1 Distinct coverage}\label{distinct-coverage}}

We now impose a search budget. Consider an arbitrary sequence of
inspected vertices \(X_1,\ldots,X_M\). The choice of \(X_t\) may depend
on the graph, on the seed, on the previously inspected vertices, and on
internal randomness. It may therefore represent a deterministic
exploration rule, a random walk, an adaptive traversal, or any other
correlated process.

The only restriction is that the rule is \textbf{target-blind}: target
identity is not available when choosing which vertex to inspect next. An
inspected vertex can of course be recognized as a target, in which case
the search succeeds and may stop. Equivalently, conditional on the
graph, the seed, and the rule's internal randomness, the counterfactual
full-budget trajectory is generated without using the target labels;
target information enters only through the stopping event. This is the
independence property required below.

If a physical search stops at its first hit, it remains useful to define
what the same rule \emph{would have done} had it continued to the full
budget \(M\). This continuation is bookkeeping only: it does not alter
whether a target was found, but it defines a full-budget inspected set
independently of target placement. Let \(A_M=\{X_1,\ldots,X_M\}\) be the
set of \textbf{distinct} vertices in that trajectory and let
\(D_M=|A_M|\). Since there are \(M\) steps, and since inspections
confined to \(B(x_0,R)\) cannot exhaust more than the ball,

\begin{equation}\label{eq:5}
D_M\le\min\{M,V(R)\}.
\end{equation}

The quantity \(D_M\) counts the genuinely different opportunities the
search has had. Repeated visits consume budget without creating new
opportunities.

\hypertarget{exact-miss-probability-and-the-universal-bound}{%
\subsection{3.2 Exact miss probability and the universal
bound}\label{exact-miss-probability-and-the-universal-bound}}

Conditional on any realized inspected set \(A_M\), the targets are still
a uniformly random \(m_q\)-subset of the graph, precisely because
\(A_M\) was constructed without target information. To miss every
target, all \(m_q\) of them must lie among the \(N-D_M\) vertices
outside \(A_M\). Hence, exactly,

\begin{equation}\label{eq:6}
\Pr(A_M\cap T_q=\varnothing\mid A_M)
=
\frac{\binom{N-D_M}{m_q}}
     {\binom{N}{m_q}}.
\end{equation}

The bound itself has a simpler reading. For a fixed \(A_M\), each of its
\(D_M\) distinct vertices has marginal target probability \(m_q/N\), so
the probability of at least one hit cannot exceed the expected number of
hits. Averaging over the search trajectory gives

\begin{equation}\label{eq:7}
P_{\rm succ}
\le
\frac{m_q}{N}\,\mathbb E[D_M]
\le
\frac{Mm_q}{N}
\equiv
\eta_q.
\end{equation}

No assumption of independent sampling, stationarity, or rapid mixing is
involved, and no property of the graph is used. Eq.~\eqref{eq:7} therefore
applies to deterministic searches, random walks, and other correlated
target-blind trajectories alike.

\hypertarget{revisits-revisit-suppression-and-structural-bias}{%
\subsection{3.3 Revisits, revisit suppression, and structural
bias}\label{revisits-revisit-suppression-and-structural-bias}}

Eq.~\eqref{eq:7} has two immediate consequences that are worth stating
separately, because they rule out the most natural attempts to escape
it.

First, \textbf{revisits act only through \(D_M\)}. For a random walk,
writing \(M_{\rm eff}\equiv\mathbb E[D_M]\) for the effective number of
distinct inspections, the sparse expansion of Eq.~\eqref{eq:6} is uniform over
the support of \(D_M\) when \(\eta_q\to0\), and

\begin{equation}\label{eq:8}
P_{\rm RW}
\sim
\frac{m_q}{N}M_{\rm eff}.
\end{equation}

Mixing times, cover times, and first-passage statistics all influence
\(D_M\), but none of them can violate Eq.~\eqref{eq:7}. Conversely, exploration
rules designed to suppress revisits --- self-avoiding or
non-backtracking walks --- keep \(D_M\) closer to \(M\) and therefore
approach the ideal distinct-inspection benchmark of Sec. 5. This uses
the finite budget better, but even in the limiting case \(D_M\simeq M\)
the scaling is unchanged.

Second, \textbf{structural bias is not relevance bias}. Target-blindness
does not require the inspection process to be structurally uniform. A
walk may preferentially visit high-degree, central, or PageRank-like
vertices, producing a strongly nonuniform visitation distribution. Such
a preference can substantially change \emph{where} the search spends its
budget, but under the neutral ensemble the target labels are assigned
independently of structural position, so preferential exposure of
structurally prominent vertices yields no systematic enrichment toward
targets. Only a preference correlated with relevance itself can help.
Detailed derivations are given in Appendix SI-1.

\hypertarget{accessibility-versus-searchability}{%
\section{4. Accessibility versus
searchability}\label{accessibility-versus-searchability}}

The significance of the finite-budget bound emerges when it is combined
with the independently defined accessibility scale. Equations (2) and
(7) are the central result:

\begin{equation}\label{eq:9}
V(R_c)\sim\frac{N}{m_q},
\qquad
P_{\rm succ}\le\eta_q=\frac{Mm_q}{N}
\sim\frac{M}{V(R_c)}.
\end{equation}

The first relation says when a target becomes reachable with finite
probability. The second says what fraction of that characteristic
accessible volume a finite target-blind search can effectively inspect.
If

\begin{equation*}
M\ll V(R_c),
\end{equation*}

or equivalently \(\eta_q\ll1\), a target can already be nearby while the
probability of finding it remains very small. Nothing in Eq.~\eqref{eq:9} assumes
hyperbolic geometry: geometry enters only through the growth of \(V(R)\)
and hence through the distance \(R_c\) at which the accessibility scale
is reached.

\textbf{Correlated targets.} The neutral ensemble is a null model, and
one may reasonably ask whether the bound is an artifact of independent
placement. It is not.

Call a target ensemble \textbf{marginally neutral} if \(|T_q|=m_q\)
almost surely and

\begin{equation*}
\Pr(u\in T_q)=\frac{m_q}{N}
\qquad\text{for every }u\in\mathcal V,
\end{equation*}

while allowing target labels at different vertices to be arbitrarily
correlated. For any such ensemble and any target-blind inspected set
\(A_M\) constructed independently of the labels, the union bound applies
unchanged:

\begin{equation*}
\Pr(A_M\cap T_q\neq\varnothing\mid A_M)
\le
|A_M|\frac{m_q}{N},
\end{equation*}

so Eq.~\eqref{eq:7} continues to hold. Clustering of relevant material therefore
does not by itself relax the finite-budget obstruction. What clustering
changes first is the accessibility probability \(\Psi(R)\) and the
fluctuation structure of \(m_R\), not the searchability ceiling. On
heterogeneous graphs, a topologically compact target set can
additionally correlate relevance with structural position; that
correlation is genuine information and falls outside the neutral model
by construction. Both effects are analyzed in Appendix SI-2.

\hypertarget{geometry-and-the-neutral-benchmark}{%
\section{5. Geometry and the neutral
benchmark}\label{geometry-and-the-neutral-benchmark}}

Geometry determines only how rapidly the accessibility scale is reached.
Inverting the ball-growth law gives

\begin{equation}\label{eq:10}
R_c
\sim
V^{-1}\left(\frac{N}{m_q}\right).
\end{equation}

For polynomial growth,

\begin{equation}\label{eq:11}
V(R)\propto R^\alpha,
\qquad
R_{c,\rm P}
\propto
\left(\frac{N}{m_q}\right)^{1/\alpha},
\end{equation}

whereas exponential graph-distance growth gives

\begin{equation}\label{eq:12}
V(R)\propto e^{\kappa R},
\qquad
R_{c,\rm E}
\propto
\frac{1}{\kappa}
\log\left(\frac{N}{m_q}\right).
\end{equation}

The hyperbolic random graph (HRG) used in Fig. 1 need not have a pure
exponential graph-distance growth law over the finite sizes studied; in
the scale-free Krioukov regime, graph distances can in fact grow more
slowly than logarithmically with \(N\) (Abdullah, Fountoulakis, and Bode
2017). We therefore measure \(V(R)\) and \(R_c^{\rm op}\) directly for
every graph and seed rather than imposing a functional form.

\hypertarget{the-exact-neutral-benchmark}{%
\subsection{5.1 The exact neutral
benchmark}\label{the-exact-neutral-benchmark}}

For comparison with Fig. 1, consider the ideal neutral
distinct-inspection benchmark within \(B(x_0,R)\): use the available
budget to inspect as many distinct accessible nodes as possible, which
by Eq.~\eqref{eq:5} means

\begin{equation*}
D(R)=\min\{M,V(R)\}
\end{equation*}

distinct vertices. Under neutral target placement, any choice of
\(D(R)\) distinct nodes has the same success probability,

\begin{equation}\label{eq:13}
\Phi_0(R)
=
1-
\frac{\binom{N-D(R)}{m_q}}
     {\binom{N}{m_q}}.
\end{equation}

Eq.~\eqref{eq:13} is an upper envelope rather than an achievable target for
every dynamics: a trajectory constrained by local motion or revisits
inspects fewer distinct vertices and performs no better. Once
\(V(R)\ge M\), increasing the radius cannot improve this benchmark at
all, so a budget-saturation scale \(V(R_B)\sim M\) is crossed at a
parametrically smaller volume than the accessibility scale whenever
\(\eta_q\ll1\). In the sparse large-\(N\) limit,

\begin{equation*}
\Phi_0\simeq1-e^{-\eta_q},
\end{equation*}

which is the asymptotic form of the exact finite-population expression,
not an independent-sampling assumption. Appendix SI-4 gives the
finite-\(N\) analysis.

Figure 1 shows the consequence. In the HRG ensemble, the target becomes
accessible at about three graph hops over the size range studied. On the
two-dimensional lattice the corresponding radius grows algebraically.
Yet the target-blind success ceiling is the same \(\eta_q=Mm_q/N\) in
both cases. Rapid expansion changes the distance at which a target
becomes reachable; it does not change the finite number of nodes that
can be inspected without information about relevance.

\begin{figure}
\centering
\includegraphics{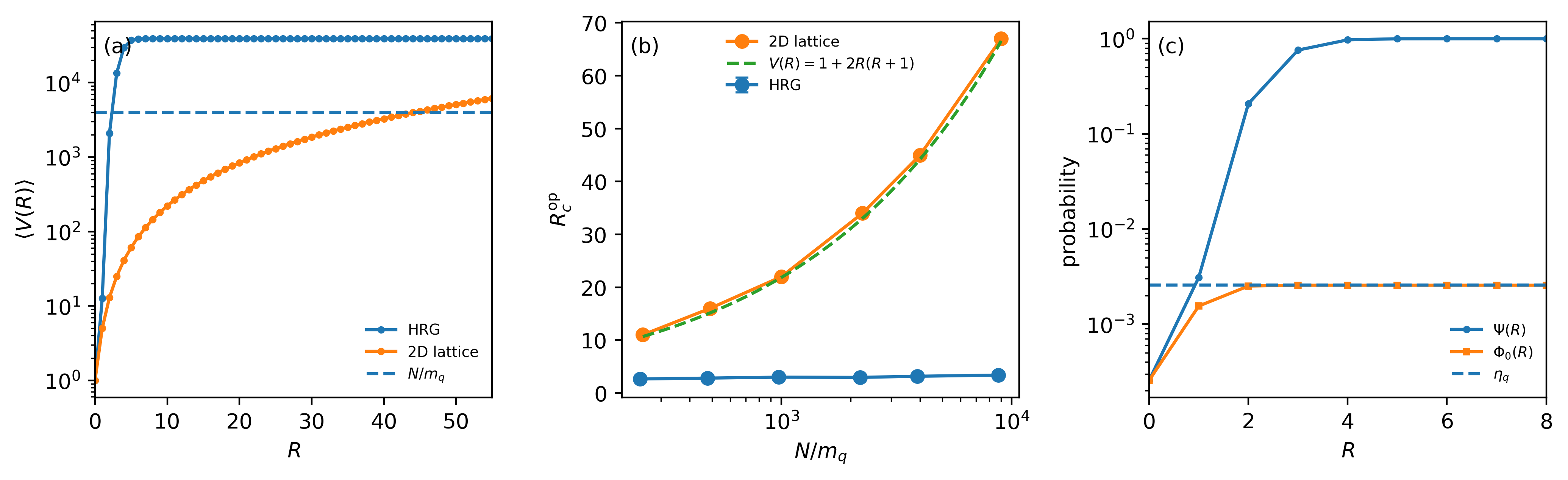}
\caption{\textbf{Geometry separates accessibility from target-blind
searchability.} (a) Mean graph-distance ball volume
\(\langle V(R)\rangle\) for Krioukov hyperbolic random graphs (HRG) and
a two-dimensional periodic lattice, for generated size
\(N_0=4\times10^4\). The dashed line is the generated-size guide
\(N_0/m_q\), with \(m_q=10\); the operational accessibility radius in
each HRG realization is instead computed using that realization's actual
giant-component size \(N\). The HRG crosses the accessibility scale
within a few graph hops, whereas the lattice grows much more gradually.
(b) Operational accessibility radius \(R_c^{\rm op}\) versus \(N/m_q\).
Error bars denote standard deviations across graph-level means after
averaging over seeds within each graph and are smaller than the symbol
size. The lattice follows the prediction from \(V(R)=1+2R(R+1)\). (c)
Accessibility probability \(\Psi(R)\) and exact neutral
distinct-inspection benchmark \(\Phi_0(R)\) for the HRG ensemble with
generated size \(N_0=4\times10^4\), with \(m_q=M=10\). Although
\(\Psi(R)\) rapidly approaches unity, \(\Phi_0(R)\) saturates at the
finite-budget ceiling \(\eta_q=Mm_q/N\simeq2.6\times10^{-3}\). Because
the HRG degree distribution is broad, some high-degree seeds already
have \(V(1)>M\), which is why the two ensemble-averaged curves separate
before the accessibility radius is reached. HRG parameters are mean
degree \(8\), degree exponent \(\gamma=2.5\), and temperature \(T=0.5\);
results use 10 graph realizations and 30 seeds per realization,
restricted to the giant connected component.}
\end{figure}

\hypertarget{what-successful-search-requires}{%
\section{6. What successful search
requires}\label{what-successful-search-requires}}

\hypertarget{relevance-enrichment}{%
\subsection{6.1 Relevance enrichment}\label{relevance-enrichment}}

Finding a target with finite probability therefore requires information
correlated with relevance. How much is required can again be read from a
counting bound.

Fix a radius \(R\) for which at least one target is present in
\(B(x_0,R)\), and suppose the search inspects only nodes in
\(B(x_0,R)\). The fraction of accessible nodes that are targets is

\begin{equation*}
\rho_R=\frac{m_R}{V(R)}.
\end{equation*}

Let \(p_R\) be the average probability, over the \(M\) inspection steps,
that the inspected node is a target. Different steps may be arbitrarily
correlated. Since success requires at least one target hit, the union
bound gives

\begin{equation}\label{eq:14}
P_{\rm succ}\le Mp_R.
\end{equation}

Thus a fixed success probability \(P_{\rm succ}\ge\delta\), with
\(0<\delta<1\), requires \(p_R\ge\delta/M\). Uniform inspection of the
accessible ball would give target probability \(\rho_R\) per step, so we
define the enrichment factor

\begin{equation*}
E_R=\frac{p_R}{\rho_R},
\end{equation*}

which measures how much more strongly the search samples nodes likely to
be relevant than it would under uniform inspection. Eq.~\eqref{eq:14} gives

\begin{equation}\label{eq:15}
E_R
\ge
\frac{\delta}{M\rho_R}.
\end{equation}

At the characteristic accessibility scale defined by Eq.~\eqref{eq:2},
\(V(R_c)\sim N/m_q\). On a discrete graph, the first shell crossing the
mean-count threshold can overshoot it substantially. Nevertheless, under
the neutral fixed-size target ensemble, at the operational
first-crossing radius,

\begin{equation*}
\rho_{R_c^{\rm op}}=\Theta_p\!\left(\frac{m_q}{N}\right)
\end{equation*}

conditional on accessibility, even when the first shell overshoots the
mean-count criterion; Appendix SI-3 proves this with explicit
probability bounds. Eq.~\eqref{eq:15} therefore yields

\begin{equation}\label{eq:16}
E_{R_c^{\rm op}}
\gtrsim
\frac{\delta N}{Mm_q}
=
\frac{\delta}{\eta_q}.
\end{equation}

In words, if target-blind search can inspect only a fraction \(\eta_q\)
of the characteristic accessible volume, then maintaining finite success
requires relevance enrichment of at least order \(1/\eta_q\) relative to
uniform inspection.

\hypertarget{candidate-sets-and-the-vanishing-visible-fraction}{%
\subsection{6.2 Candidate sets and the vanishing visible
fraction}\label{candidate-sets-and-the-vanishing-visible-fraction}}

Many practical search, ranking, and recommendation systems do not assign
a smoothly varying inspection probability to every accessible node. They
restrict attention to a much smaller set of candidates and then rank or
inspect within it. Let \(S_R\subseteq B(x_0,R)\) be such a
\textbf{candidate set}; retaining the edges between selected vertices
makes it an induced candidate or visible subgraph, though no graph
structure is required for what follows.

Suppose inspection is uniform within \(S_R\), and define the visible
fraction \(\phi_R=|S_R|/V(R)\), the local recall
\(a_R=|S_R\cap T_q|/m_R\), and the target fraction inside the candidate
set \(P_R=|S_R\cap T_q|/|S_R|\). These obey the identity
\(P_R=(a_R/\phi_R)\rho_R\), so the enrichment takes the transparent form

\begin{equation}\label{eq:17}
E_R=\frac{P_R}{\rho_R}=\frac{a_R}{\phi_R}.
\end{equation}

To raise target density above the accessible baseline, an algorithm must
retain targets more efficiently than it retains ordinary accessible
nodes. Applying the union bound to uniform inspection within \(S_R\)
requires \(P_R\ge\delta/M\), hence
\(\phi_R\le(M/\delta)a_R\rho_R\le(M/\delta)\rho_R\). Using the density
result of Sec. 6.1,

\begin{equation}\label{eq:18}
\phi_{R_c^{\rm op}}
=
O_p\!\left(\frac{\eta_q}{\delta}\right)
\qquad\text{conditional on accessibility.}
\end{equation}

As \(\eta_q\to0\), any uniformly searched candidate set that maintains a
fixed success probability must therefore occupy a vanishing fraction of
the accessible region. For such implementations, the chain

\begin{equation*}
\text{accessible region}
\;\longrightarrow\;
\text{relevance-selected candidate set}
\;\longrightarrow\;
\text{inspection}
\end{equation*}

expresses the required contraction of the visible region. More
generally, successful search requires a comparable concentration of
inspection probability toward relevance, whether or not it constructs an
explicit candidate set. Appendix SI-5 gives the exact finite-\(N\)
version, the associated recall bound, and the precision--recall
geometry.

\hypertarget{discussion}{%
\section{7. Discussion}\label{discussion}}

The result can be summarized by one ratio. At the accessibility radius
the reachable region contains of order \(V(R_c)\sim N/m_q\) nodes, while
the search has only \(M\) inspection steps, so their ratio is
\(M/V(R_c)\sim\eta_q\). This is why rapid reachability and poor
searchability can coexist: geometry can compress graph distance without
increasing the inspection budget. This complements the
decentralized-routing perspective: structural cues can make a known
destination navigable, but search for an unknown relevant target
requires cues that correlate specifically with relevance.

Real search engines, ranking systems, and recommendation mechanisms are
not target-blind. The result does not say that algorithms cannot make
search effective. It identifies what successful search must supply in
the sparse regime: information that shifts inspection probability toward
nodes more likely to be relevant. Eq.~\eqref{eq:16} quantifies the minimum
enrichment required, and Eq.~\eqref{eq:18} gives its geometric counterpart for
uniformly searched candidate sets: a vanishing visible fraction. Ranking
and recommendation provide familiar examples of strongly nonuniform
exposure (Joachims et al. 2007; Huszár et al. 2022; Fabbri et al. 2022).

The same counting logic applies beyond graph-distance geometry. For
retrieval in hyperbolic embedding spaces (Nickel and Kiela 2017; Ganea,
Bécigneul, and Hofmann 2018; Prokhorenkova et al. 2022), the argument
goes through once a candidate region and a finite inspection budget are
specified; what changes is that the relevant geometry is that of the
representation and of the candidate-generation procedure rather than of
the original network.

More broadly, the result gives a precise finite-budget form to a simple
consequence of rapid growth of accessible configuration space:
exploration that remains neutral with respect to relevance quickly
becomes ineffective, whereas useful search must select among the
proliferating accessible alternatives. When only a fraction \(\eta_q\)
of the characteristic accessible volume can be inspected, nonvanishing
success requires a relevance enrichment of order \(1/\eta_q\) relative
to uniform inspection. In this sense, the relevant question is not
whether effective search must be selective, but which relevance signal
guides that selectivity.

For scale, consider \(N\sim10^{10}\) nodes, \(m_q\sim10^3\) relevant
nodes, and \(M\sim10\) inspection steps. Then

\begin{equation*}
\eta_q\sim10^{-6}.
\end{equation*}

A target can become geometrically accessible while neutral search
remains limited to a success probability of order one in a million.
Achieving order-one success instead requires relevance enrichment of
order \(10^6\) relative to uniform inspection. Which relevance signal
produces that enrichment is algorithm-dependent; the required scale of
enrichment is not.

The supplementary material provides the detailed treatment of correlated
target-blind exploration and revisits, clustered target ensembles,
fluctuations at the discrete accessibility radius, exact finite-budget
results, candidate-set consequences, and numerical implementation.

\hypertarget{acknowledgments}{%
\section{Acknowledgments}\label{acknowledgments}}

Generative AI tools (OpenAI ChatGPT and Anthropic Claude) were used
substantively to assist with manuscript English language editing, grammar 
checking, stylistic improvements, proofreading and code debugging. The author directed
the use of these tools, reviewed and verified their outputs, and remains responsible for 
all derivations, numerical procedures, citations, and scientific claims.

\hypertarget{data-availability}{%
\section{Data Availability}\label{data-availability}}

The data and code that support the findings of this study are openly available in a Zenodo repository at \url{https://doi.org/10.5281/zenodo.22955225}. 
The deposit contains the archived numerical output at graph/seed/radius resolution, the generation and plotting scripts, and a verification
harness that checks the archived inputs against reference hashes and reproduces every numerical series plotted in Fig. 1.

\appendix

\hypertarget{supplementary-material}{%
\section{Supplementary material}\label{supplementary-material}}

\emph{Derivations, robustness analyses, candidate-set bounds, and
numerical implementation supporting the article.}

\hypertarget{purpose-and-roadmap}{%
\section{Purpose and roadmap}\label{purpose-and-roadmap}}

The main text separates two questions that are often conflated.

First, \textbf{accessibility} asks when a rare target is likely to lie
inside the region that can be reached from a seed. If the fraction of
target nodes is \(m_q/N\) and the graph-distance ball contains \(V(R)\)
nodes, the mean number of accessible targets is

\begin{equation*}
\lambda_R=\frac{m_qV(R)}{N}.
\end{equation*}

The characteristic accessibility scale is therefore

\begin{equation*}
V(R_c)\sim\frac{N}{m_q}.
\end{equation*}

Second, \textbf{searchability} asks whether a finite search can actually
find a target once it is accessible. A target-blind search that has at
most \(M\) inspection steps satisfies

\begin{equation*}
P_{\mathrm{succ}}\le\frac{Mm_q}{N}\equiv\eta_q.
\end{equation*}

At the characteristic accessibility scale,

\begin{equation*}
\eta_q\sim\frac{M}{V(R_c)}.
\end{equation*}

Thus a target may already be nearby while only a very small fraction of
the characteristic accessible region can be inspected.

The supplementary material gives the details behind these statements and
clarifies their scope:

\begin{itemize}
\tightlist
\item
  \textbf{SI-1} proves the target-blind bound for arbitrary correlated
  search trajectories and shows how revisits reduce effective coverage.
\item
  \textbf{SI-2} explains what changes when target locations are
  clustered rather than independently or uniformly scattered.
\item
  \textbf{SI-3} treats fluctuations in the number of accessible targets
  and the discrete first-crossing radius used on finite graphs.
\item
  \textbf{SI-4} gives the exact finite-\(N\) neutral benchmark shown in
  Fig. 1.
\item
  \textbf{SI-5} derives the relevance-enrichment requirement and makes
  explicit the candidate-set or visible-subgraph interpretation used in
  the main text.
\item
  \textbf{SI-6} describes the numerical implementation for Fig. 1.
\end{itemize}

\hypertarget{conventions-and-notation}{%
\section{Conventions and notation}\label{conventions-and-notation}}

The graph is denoted as

\begin{equation*}
G=(\mathcal V,E),
\qquad
|\mathcal V|=N,
\end{equation*}

where \(\mathcal V\) is the vertex set and \(E\) the edge set. We fix a
seed node \(x_0\). The graph-distance ball of radius \(R\) is

\begin{equation*}
B(x_0,R)=\{v\in\mathcal V:d(x_0,v)\le R\},
\end{equation*}

and its number of vertices is

\begin{equation*}
V_R\equiv V(R)=|B(x_0,R)|.
\end{equation*}

The notation \(V_R\) is used in the supplementary material when it
shortens formulas; it is the same quantity denoted \(V(R)\) in the main
text.

For a target class labeled by \(q\), let

\begin{equation*}
T_q\subseteq\mathcal V,
\qquad
|T_q|=m_q,
\end{equation*}

be the set of relevant vertices. The \textbf{neutral fixed-size target
ensemble} means that \(T_q\) is chosen uniformly among all subsets of
\(\mathcal V\) containing exactly \(m_q\) vertices. Hence every vertex
has the same marginal probability

\begin{equation*}
\Pr(v\in T_q)=\frac{m_q}{N}.
\end{equation*}

For a fixed target placement, the number of targets inside the
accessible ball and their local density are

\begin{equation*}
m_R=|T_q\cap B(x_0,R)|,
\qquad
\rho_R=\frac{m_R}{V_R}.
\end{equation*}

A search has at most \(M\) inspection steps. The dimensionless
finite-budget parameter is

\begin{equation*}
\eta_q=\frac{Mm_q}{N}.
\end{equation*}

The regime \(\eta_q\ll1\) is the sparse regime in which a target-blind
search can inspect only a small fraction of the characteristic
accessibility volume. When we write \(\eta_q\to0\), we mean simply that
this inspectable fraction tends to zero.

We say that a search has \textbf{finite success} if

\begin{equation*}
P_{\mathrm{succ}}\ge\delta,
\qquad
0<\delta<1,
\end{equation*}

where \(\delta\) is held fixed as the system size changes.

An average over random target placements is denoted by \(\mathbb E_T\).
When graphs and seed vertices are also sampled, we use \(\mathbb E_G\)
for that geometric average. A statement for one fixed graph, seed, and
target placement is sometimes called a \textbf{fixed-realization} or
\textbf{quenched} statement. A statement after averaging over target
placements is sometimes called \textbf{target-averaged} or
\textbf{annealed}.

For a sequence of non-negative random variables, \(X_N=O_p(a_N)\) means
that \(X_N/a_N\) remains bounded with arbitrarily high probability as
\(N\) increases. We write \(X_N=\Theta_p(a_N)\) when \(X_N/a_N\) is
bounded both above and away from zero with arbitrarily high probability;
equivalently, \(X_N=O_p(a_N)\) and \(a_N=O_p(X_N)\).

The characteristic accessibility radius \(R_c\) is defined at the level
of scaling by

\begin{equation*}
V(R_c)\sim\frac{N}{m_q},
\end{equation*}

or equivalently by a mean target count of order one,

\begin{equation*}
\lambda_{R_c}
=
\frac{m_qV(R_c)}{N}
\sim1.
\end{equation*}

On an actual discrete graph, \(R\) takes integer values and \(V(R)\) can
jump strongly from one shell to the next. We therefore also use the
\textbf{operational first-crossing radius}

\begin{equation*}
R_c^{\mathrm{op}}
=
\min\left\{R:\frac{m_qV_R}{N}\ge1\right\}.
\end{equation*}

The two definitions answer slightly different questions. \(R_c\) is the
characteristic scale obtained from the growth law; \(R_c^{\mathrm{op}}\)
is the first integer graph distance at which the measured ball crosses
the criterion \(\lambda_R=1\). SI-3 shows that the main
relevance-enrichment scaling remains valid even if the first crossing
overshoots this criterion substantially.

\hypertarget{si-1-correlated-target-blind-search-distinct-coverage-and-revisits}{%
\section{SI-1: Correlated target-blind search -- distinct coverage and
revisits}\label{si-1-correlated-target-blind-search-distinct-coverage-and-revisits}}

\hypertarget{si-1.1-what-target-blind-means}{%
\subsection{SI-1.1 What ``target-blind''
means}\label{si-1.1-what-target-blind-means}}

Consider an arbitrary sequence of inspected vertices

\begin{equation*}
X_1,X_2,\ldots,X_M.
\end{equation*}

The choice of \(X_t\) may depend on the graph, the seed \(x_0\), the
previous inspected vertices, and internal randomness. It may therefore
represent a deterministic exploration rule, a random walk, an adaptive
graph traversal, or any other correlated process.

The only restriction is that the rule is target-blind: target identity
is not available when choosing which vertex to inspect next. An
inspected vertex can of course be recognized as a target, in which case
the search succeeds and may stop. Equivalently, conditional on the
graph, the seed, and the rule's internal randomness, the counterfactual
full-budget trajectory is generated without using the target labels;
target information enters only through the stopping event when a target
is encountered. This is the independence property needed in the counting
argument below.

If a physical search stops when it first finds a target, it is still
useful mathematically to define what the same target-blind rule
\emph{would have done} had it continued to the full budget \(M\). This
continuation is only bookkeeping. It does not alter whether a target was
found during the actual search, but it lets us define a full-budget
inspected set independently of target placement.

Let

\begin{equation*}
A_M=\{X_1,\ldots,X_M\}
\end{equation*}

be the set of \textbf{distinct} vertices appearing in that full-budget
trajectory, and let

\begin{equation*}
D_M=|A_M|.
\end{equation*}

Because there are only \(M\) inspection steps,

\begin{equation*}
D_M\le M.
\end{equation*}

If all inspected vertices are restricted to \(B(x_0,R)\), then also

\begin{equation*}
D_M\le V_R.
\end{equation*}

Therefore

\begin{equation}
D_M\le\min\{M,V_R\}.
\tag{S1.1}
\end{equation}

The quantity \(D_M\) is the number of genuinely different opportunities
the search has had. Repeated visits consume budget but do not create new
target opportunities.

\hypertarget{si-1.2-exact-success-probability-for-a-given-inspected-set}{%
\subsection{SI-1.2 Exact success probability for a given inspected
set}\label{si-1.2-exact-success-probability-for-a-given-inspected-set}}

Fix a realized target-blind set \(A_M\) containing \(d\) distinct
vertices. Under the neutral target ensemble, the \(m_q\) targets are
still a uniformly random \(m_q\)-subset of the \(N\) graph vertices,
because the construction of \(A_M\) did not use target information.

To miss every target, all \(m_q\) targets must lie among the \(N-d\)
vertices outside \(A_M\). Hence

\begin{equation}
\Pr(A_M\cap T_q=\varnothing\mid A_M)
=
\frac{\binom{N-d}{m_q}}{\binom{N}{m_q}}.
\tag{S1.2}
\end{equation}

The same probability can be written by reversing the counting: choose
the \(d\) inspected vertices from the \(N-m_q\) non-target vertices,

\begin{equation}
\Pr(A_M\cap T_q=\varnothing\mid D_M=d)
=
\frac{\binom{N-m_q}{d}}{\binom{N}{d}}.
\tag{S1.3}
\end{equation}

Define

\begin{equation}
F(d)
=
1-
\frac{\binom{N-d}{m_q}}{\binom{N}{m_q}}.
\tag{S1.4}
\end{equation}

Then \(F(d)\) is the exact probability that a neutral target set
intersects any fixed set of \(d\) vertices. Averaging over the random
search trajectory gives

\begin{equation}
P_{\mathrm{succ}}=\mathbb E[F(D_M)].
\tag{S1.5}
\end{equation}

This equation is useful because it isolates the role of correlations in
the search trajectory. Under neutral target placement, the \emph{shape}
of the target-blind inspected set is irrelevant once its cardinality
\(D_M\) is known. Geometry and dynamics can influence how large \(D_M\)
becomes, but they do not make one target-blind set of size \(d\) more
relevant than another.

\hypertarget{si-1.3-universal-finite-budget-bound}{%
\subsection{SI-1.3 Universal finite-budget
bound}\label{si-1.3-universal-finite-budget-bound}}

A simpler argument yields the universal finite-budget bound directly.

For a fixed inspected set of \(d\) vertices, each inspected vertex has
marginal target probability \(m_q/N\). The expected number of target
vertices in the inspected set is therefore

\begin{equation*}
d\frac{m_q}{N}.
\end{equation*}

The probability of at least one hit cannot exceed the expected number of
hits. Equivalently, this is the union bound. Thus

\begin{equation*}
F(d)\le d\frac{m_q}{N}.
\end{equation*}

Using Eq. (S1.5),

\begin{equation}
P_{\mathrm{succ}}
\le
\frac{m_q}{N}\mathbb E[D_M]
\le
\frac{Mm_q}{N}
=
\eta_q.
\tag{S1.6}
\end{equation}

No independence between inspection steps is assumed. No mixing time is
assumed. No particular graph geometry is assumed.

If

\begin{equation*}
d=o(N),
\qquad
m_q=o(N),
\qquad
\frac{d\,m_q}{N}\to0,
\end{equation*}

then the exact hypergeometric probability has the sparse expansion

\begin{equation}
F(d)\sim\frac{d\,m_q}{N}.
\tag{S1.7}
\end{equation}

In that regime, target-blind performance is therefore controlled to
leading order by the number of distinct vertices inspected.

\hypertarget{si-1.4-random-walks-revisit-suppression-and-structural-bias}{%
\subsection{SI-1.4 Random walks, revisit suppression, and structural
bias}\label{si-1.4-random-walks-revisit-suppression-and-structural-bias}}

A simple random walk is only one example of a correlated target-blind
search. Let \(D_M\) denote the number of distinct vertices visited
during the first \(M\) inspection steps. Its exact neutral success
probability is

\begin{equation}
P_{\mathrm{RW}}(M)
=
\mathbb E\left[
1-\frac{\binom{N-D_M}{m_q}}{\binom{N}{m_q}}
\right].
\tag{S1.8}
\end{equation}

If the walk revisits vertices, then typically \(D_M<M\). A convenient
measure of the effective number of distinct inspection opportunities is
therefore

\begin{equation}
M_{\mathrm{eff}}
\equiv
\mathbb E[D_M].
\tag{S1.9}
\end{equation}

Because \(D_M\le M\), the condition

\begin{equation*}
\eta_q=\frac{Mm_q}{N}\to0
\end{equation*}

makes the sparse expansion uniform over the entire support of \(D_M\).
More precisely, for every \(0\le d\le M\),

\begin{equation*}
F(d)
=
\frac{d\,m_q}{N}
\left[1+O(\eta_q)\right].
\end{equation*}

Therefore

\begin{equation}
P_{\mathrm{RW}}
=
\frac{m_q}{N}\mathbb E[D_M]
\left[1+O(\eta_q)\right]
=
\frac{m_q}{N}M_{\mathrm{eff}}
\left[1+O(\eta_q)\right].
\tag{S1.10}
\end{equation}

Hence, as \(\eta_q\to0\),

\begin{equation*}
P_{\mathrm{RW}}
\sim
\frac{m_q}{N}M_{\mathrm{eff}}.
\end{equation*}

Thus revisits reduce target-blind performance only by reducing the
number of distinct vertices inspected. Mixing properties, cover times,
and first-passage statistics can all affect \(D_M\), but none of them
can violate the universal bound in Eq. (S1.6).

The same framework also covers exploration rules designed explicitly to
suppress revisits. Self-avoiding or non-backtracking walks can keep
\(D_M\) much closer to \(M\), and therefore approach the ideal
distinct-inspection benchmark. This improves the use of the finite
budget, but it does not change the scaling of neutral searchability:
even in the limiting case \(D_M\simeq M\),

\begin{equation*}
P_{\mathrm{succ}}
\lesssim
\frac{Mm_q}{N}.
\end{equation*}

Target-blindness also does not require the inspection process to be
structurally uniform. A walk may preferentially visit high-degree,
central, or PageRank-like vertices, producing a strongly nonuniform
stationary or visitation distribution. Such a structural bias can
substantially change where the search spends its budget, but it is not
by itself a relevance bias. Under the neutral target ensemble, target
labels are assigned independently of that structural preference, so
preferential exposure of structurally prominent vertices provides no
systematic enrichment toward targets.

These examples separate three distinct effects. Ordinary random walks
may waste budget through revisits; self-avoiding or non-backtracking
exploration can reduce that waste; structurally biased walks can
redistribute inspection probability across the graph. All remain subject
to Eq. (S1.6) as long as the rule used to choose where to inspect is
independent of the target labels.

\hypertarget{si-1.5-what-the-bound-says}{%
\subsection{SI-1.5 What the bound
says}\label{si-1.5-what-the-bound-says}}

Combining Eq. (S1.1) with the monotonicity of \(F(d)\) gives

\begin{equation}
P_{\mathrm{succ}}
=
\mathbb E[F(D_M)]
\le
F(\min\{M,V_R\})
\le
\eta_q.
\tag{S1.11}
\end{equation}

The first inequality says that the best target-blind use of a finite
accessible region is to inspect as many \emph{distinct} vertices as
possible. The second says that even perfect distinct coverage cannot
overcome the global finite-budget factor \(Mm_q/N\).

The exact benchmark \(F(\min\{M,V_R\})\) is analyzed in SI-4.

\hypertarget{si-2-clustered-targets-what-changes-and-what-does-not}{%
\section{SI-2: Clustered targets -- what changes and what does
not}\label{si-2-clustered-targets-what-changes-and-what-does-not}}

The neutral ensemble used in the main text deliberately removes
correlations between relevance and graph structure. Real targets can
instead be clustered. This section separates two distinct effects:

\begin{enumerate}
\def\labelenumi{\arabic{enumi}.}
\tightlist
\item
  \textbf{target--target correlation}, meaning that nearby vertices tend
  to be relevant together;
\item
  \textbf{structural relevance bias}, meaning that some graph positions
  are intrinsically more likely to contain targets than others.
\end{enumerate}

The two effects are not equivalent.

\hypertarget{si-2.1-correlated-target-ensembles-with-uniform-one-vertex-marginals}{%
\subsection{SI-2.1 Correlated target ensembles with uniform one-vertex
marginals}\label{si-2.1-correlated-target-ensembles-with-uniform-one-vertex-marginals}}

The neutral fixed-size ensemble has

\begin{equation*}
\Pr(u\in T_q)=\frac{m_q}{N}
\qquad\text{for every }u\in\mathcal V.
\end{equation*}

Independence of different target indicators is not required for this
property. We call a correlated target ensemble \textbf{marginally
neutral} if

\begin{equation*}
|T_q|=m_q
\end{equation*}

almost surely and

\begin{equation}
\Pr(u\in T_q)=\frac{m_q}{N}
\qquad\text{for every }u\in\mathcal V,
\tag{S2.1}
\end{equation}

even though target labels at different vertices may be correlated.

Let

\begin{equation*}
Y_u=\mathbf 1_{\{u\in T_q\}}.
\end{equation*}

Then

\begin{equation*}
m_R=\sum_{u\in B(x_0,R)}Y_u,
\end{equation*}

so every marginally neutral ensemble satisfies

\begin{equation}
\mathbb E_T[m_R]
=
\frac{m_q}{N}V_R,
\tag{S2.2}
\end{equation}

and

\begin{equation}
\mathbb E_T[\rho_R]
=
\frac{m_q}{N}.
\tag{S2.3}
\end{equation}

Thus clustering can leave the \textbf{mean} number of accessible targets
unchanged.

It can nevertheless change the fluctuations. In full,

\begin{gather}
\operatorname{Var}(m_R)
=
\sum_{u\in B(x_0,R)}\operatorname{Var}(Y_u)
+
2\sum_{\substack{u<v\\u,v\in B(x_0,R)}}
\operatorname{Cov}(Y_u,Y_v).
\tag{S2.4}
\end{gather}

The covariance terms vanish when the corresponding target indicators are
pairwise uncorrelated; independence is sufficient but not necessary.
Positive spatial correlations can increase the probability of both
unusually empty and unusually target-rich balls, even when the mean
occupancy is unchanged.

The mean alone therefore does not determine the accessibility
probability

\begin{equation*}
\Psi(R)=\Pr(m_R\ge1).
\end{equation*}

Since \(0\le m_R\le m_q\),

\begin{equation}
\frac{\mathbb E[m_R]}{m_q}
\le
\Psi(R)
\le
\min\{1,\mathbb E[m_R]\}.
\tag{S2.5}
\end{equation}

Thus the first moment alone does not determine the first-contact
probability; the compact-cluster construction below provides an explicit
example.

There is also a useful robustness result for target-blind search. For
any marginally neutral target ensemble and any target-blind inspected
set \(A_M\) independent of target labels,

\begin{equation*}
\Pr(A_M\cap T_q\neq\varnothing\mid A_M)
\le
|A_M|\frac{m_q}{N}.
\end{equation*}

Therefore the universal bound

\begin{equation*}
P_{\mathrm{succ}}\le\frac{Mm_q}{N}
\end{equation*}

continues to hold even when target labels are correlated, provided their
one-vertex marginals remain uniform. What clustering changes first is
accessibility and fluctuation structure, not this union-bound
obstruction.

\hypertarget{si-2.2-a-compact-target-cluster-on-a-homogeneous-graph}{%
\subsection{SI-2.2 A compact target cluster on a homogeneous
graph}\label{si-2.2-a-compact-target-cluster-on-a-homogeneous-graph}}

To see the distinction explicitly, consider a finite vertex-transitive
graph. ``Vertex-transitive'' means that all vertices have the same
graph-distance environment, so the ball size \(|B(x,r)|\) depends on
\(r\) but not on the center \(x\).

Let the target set be a compact ball

\begin{equation*}
T_q=B(x^*,r_c),
\end{equation*}

where the center \(x^*\) is chosen uniformly among graph vertices.
Vertex transitivity implies

\begin{equation*}
|B(x,r_c)|=V(r_c)
\end{equation*}

for every center, so the number of targets is fixed,

\begin{equation*}
m_q=V(r_c).
\end{equation*}

For any vertex \(u\),

\begin{equation*}
u\in B(x^*,r_c)
\quad\Longleftrightarrow\quad
x^*\in B(u,r_c).
\end{equation*}

Since \(x^*\) is uniform,

\begin{equation}
\Pr_{x^*}(u\in T_q)
=
\frac{V(r_c)}{N}
=
\frac{m_q}{N}.
\tag{S2.6}
\end{equation}

Thus this compact cluster is marginally neutral even though the target
labels are strongly correlated over short distances. Consequently,

\begin{equation}
\mathbb E_{x^*}[m_R]
=
\frac{m_q}{N}V_R.
\tag{S2.7}
\end{equation}

The first-contact probability, however, is different.

The balls \(B(x_0,R)\) and \(B(x^*,r_c)\) intersect if and only if

\begin{equation*}
d(x_0,x^*)\le R+r_c.
\end{equation*}

This statement is exact for the usual shortest-path metric on any
undirected graph: if the center-to-center distance is at most \(R+r_c\),
a vertex on a shortest path lies within both balls; if the distance is
larger, the triangle inequality forbids an intersection.

On a vertex-transitive graph the number of possible cluster centers
satisfying this condition is \(V(R+r_c)\). Therefore, exactly,

\begin{equation}
\Psi_{\mathrm{cl}}(R)
=
\frac{V(R+r_c)}{N}.
\tag{S2.8}
\end{equation}

Finite-size saturation is already included because \(V(R+r_c)=N\) once
the radius reaches the graph diameter.

For comparison, the neutral fixed-size ensemble gives

\begin{equation}
\Psi_0(R)
=
1-
\frac{\binom{N-V_R}{m_q}}{\binom{N}{m_q}}.
\tag{S2.9}
\end{equation}

Thus target clustering can preserve the average occupancy while changing
the probability that the accessible ball is empty.

If the graph has approximately multiplicative exponential ball growth,

\begin{equation*}
V(R)\asymp e^{\kappa R},
\end{equation*}

then

\begin{equation}
V(R+r_c)
\asymp
V(R)V(r_c)
\asymp
m_qV(R),
\tag{S2.10}
\end{equation}

and the clustered first-contact condition is again, to leading order,

\begin{equation}
V(R)\asymp\frac{N}{m_q}.
\tag{S2.11}
\end{equation}

This agreement is a consequence of the growth law. It is not a general
theorem saying that clustering never changes the accessibility scale.

\hypertarget{si-2.3-fixed-realization-overlap-on-a-regular-tree}{%
\subsection{SI-2.3 Fixed-realization overlap on a regular
tree}\label{si-2.3-fixed-realization-overlap-on-a-regular-tree}}

The previous subsection averaged over the random cluster center. To see
what one fixed target cluster looks like geometrically, consider a
regular tree with root degree \(b\) and branching number \(b-1\). Its
ball volume is

\begin{equation}
V_b(R)
=
1+b\frac{(b-1)^R-1}{b-2},
\qquad b>2.
\tag{S2.12}
\end{equation}

Let

\begin{equation*}
T_q=B(x^*,r_c),
\qquad
d^*=d(x_0,x^*),
\end{equation*}

and define the overlap size

\begin{equation}
J(R,d^*,r_c)
=
|B(x_0,R)\cap B(x^*,r_c)|.
\tag{S2.13}
\end{equation}

There are four elementary regimes.

\textbf{No overlap.} If

\begin{equation*}
d^*>R+r_c,
\end{equation*}

then

\begin{equation}
J(R,d^*,r_c)=0.
\tag{S2.14}
\end{equation}

\textbf{The accessible ball lies entirely inside the target cluster.} If

\begin{equation*}
d^*+R\le r_c,
\end{equation*}

then

\begin{equation}
J(R,d^*,r_c)=V_b(R),
\tag{S2.15}
\end{equation}

and therefore every accessible vertex is a target,

\begin{equation}
\rho_R=1.
\tag{S2.16}
\end{equation}

\textbf{The target cluster lies entirely inside the accessible ball.} If

\begin{equation*}
d^*+r_c\le R,
\end{equation*}

then

\begin{equation}
J(R,d^*,r_c)=V_b(r_c)=m_q.
\tag{S2.17}
\end{equation}

\textbf{Partial overlap.} In the remaining intersecting regime,

\begin{equation*}
|R-r_c|<d^*\le R+r_c,
\end{equation*}

one has

\begin{equation}
0<J(R,d^*,r_c)<\min\{V_b(R),m_q\}.
\tag{S2.18}
\end{equation}

For this fixed target realization,

\begin{equation}
\rho_R
=
\frac{J(R,d^*,r_c)}{V_b(R)}.
\tag{S2.19}
\end{equation}

The first radius at which the accessible ball touches the target cluster
is

\begin{equation}
R_{\mathrm{first}}
=
\max\{0,d^*-r_c\},
\tag{S2.20}
\end{equation}

while the radius at which the whole target cluster is contained in the
accessible ball is

\begin{equation}
R_{\mathrm{full}}
=
d^*+r_c.
\tag{S2.21}
\end{equation}

These relations are purely geometric and illustrate why a fixed
clustered target realization can have first-contact behavior very
different from the neutral ensemble even when ensemble averages agree.

\hypertarget{si-2.4-heterogeneous-graphs-clustering-can-also-bias-relevance-toward-structure}{%
\subsection{SI-2.4 Heterogeneous graphs: clustering can also bias
relevance toward
structure}\label{si-2.4-heterogeneous-graphs-clustering-can-also-bias-relevance-toward-structure}}

The homogeneous construction above does not automatically remain
marginally neutral on a heterogeneous graph.

Suppose again that the targets are constructed as a ball around a
uniformly chosen center,

\begin{equation*}
T_q=B(x^*,r_c).
\end{equation*}

For a fixed vertex \(u\),

\begin{equation}
\Pr(u\in T_q)
=
\frac{|B(u,r_c)|}{N}.
\tag{S2.22}
\end{equation}

If ball sizes depend on \(u\), then some vertices are more likely than
others to belong to a randomly centered target cluster. The target-set
size \(|B(x^*,r_c)|\) can also depend on the chosen center.

Thus this construction combines two effects:

\begin{itemize}
\tightlist
\item
  nearby target labels are correlated;
\item
  relevance itself becomes correlated with structural position.
\end{itemize}

If every realization is required to contain exactly \(m_q\) vertices, we
can use a fixed-cardinality construction. Choose a center \(x^*\)
uniformly at random and let \(r^*\) be the smallest radius for which

\begin{equation*}
|B(x^*,r^*)| \ge m_q.
\end{equation*}

Define

\begin{equation*}
T_q=C_{m_q}(x^*)
\end{equation*}

by including all vertices at distance smaller than \(r^*\), together
with a random selection of vertices at distance \(r^*\) sufficient to
make

\begin{equation*}
|T_q|=|C_{m_q}(x^*)|=m_q.
\end{equation*}

The marginal inclusion probability

\begin{equation}
q_u
=
\Pr[u\in T_q]
=
\Pr[u\in C_{m_q}(x^*)]
\tag{S2.23}
\end{equation}

where the probability is over both the random choice of \(x^*\) and the
random selection at the outer shell, need not equal \(m_q/N\).

For a fixed graph and seed,

\begin{equation}
\mathbb E[m_R]
=
\sum_{u\in B(x_0,R)}q_u.
\tag{S2.24}
\end{equation}

A convenient dimensionless measure of this structural bias is thus

\begin{equation}
\mathcal B(R;x_0)
=
\frac{\displaystyle\sum_{u\in B(x_0,R)}q_u}
     {\displaystyle(m_q/N)V_R}.
\tag{S2.25}
\end{equation}

If \(\mathcal B=1\), the mean occupancy agrees with the marginally
neutral baseline. Values above or below one indicate that the accessible
ball is, on average, enriched or depleted in relevance because of its
structural location.

This distinction matters for heterogeneous networks such as HRGs, where
graph-distance neighborhoods vary strongly with degree and radial
position.

\hypertarget{si-2.5-interpretation}{%
\subsection{SI-2.5 Interpretation}\label{si-2.5-interpretation}}

The calculations above are not needed for Fig. 1, which uses the neutral
fixed-size target ensemble. Their role is to clarify how the distinction
between accessibility and searchability should be interpreted when
relevance is structured.

On a homogeneous graph such as a periodic lattice, a compact target
cluster can change first-contact fluctuations while preserving

\begin{equation*}
\mathbb E[m_R]
=
\frac{m_q}{N}V_R.
\end{equation*}

On a heterogeneous HRG, a topologically compact target set can
additionally introduce structural relevance bias. In that case, the mean
occupancy \(\mathbb E[m_R]\), the accessibility probability
\(\Pr(m_R>0)\), the structural-bias factor \(\mathcal B(R;x_0)\), and
the distribution of first-contact radii describe distinct aspects of
target accessibility and need not convey the same information.

These distinctions do not affect the neutral benchmark studied in the
main text; they only specify how the framework should be interpreted
when target relevance is correlated with network structure.

\hypertarget{si-3-target-count-fluctuations-and-the-discrete-accessibility-radius}{%
\section{SI-3: Target-count fluctuations and the discrete accessibility
radius}\label{si-3-target-count-fluctuations-and-the-discrete-accessibility-radius}}

\hypertarget{si-3.1-exact-distribution-of-the-number-of-accessible-targets}{%
\subsection{SI-3.1 Exact distribution of the number of accessible
targets}\label{si-3.1-exact-distribution-of-the-number-of-accessible-targets}}

Fix the graph, the seed, and the radius \(R\). Under the neutral
fixed-size target ensemble, drawing the \(m_q\) target vertices is
equivalent to drawing \(m_q\) marked objects from a population of \(N\)
vertices, of which \(V_R\) lie inside the ball.

Therefore

\begin{equation}
m_R
\sim
\operatorname{Hypergeom}(N,m_q,V_R).
\tag{S3.1}
\end{equation}

Its mean is

\begin{equation}
\lambda_R
\equiv
\bar m_R
=
\mathbb E_T[m_R]
=
\frac{m_qV_R}{N},
\tag{S3.2}
\end{equation}

and its variance is

\begin{equation}
\operatorname{Var}_T(m_R)
=
V_R\frac{m_q}{N}
\left(1-\frac{m_q}{N}\right)
\frac{N-V_R}{N-1}.
\tag{S3.3}
\end{equation}

Equivalently,

\begin{equation}
\operatorname{Var}_T(m_R)
=
\lambda_R
\left(1-\frac{m_q}{N}\right)
\frac{N-V_R}{N-1}
\le
\lambda_R.
\tag{S3.4}
\end{equation}

For the local target density \(\rho_R=m_R/V_R\),

\begin{equation}
\frac{\operatorname{Var}_T(\rho_R)}
     {\mathbb E_T[\rho_R]^2}
=
\frac{1}{\lambda_R}
\left(1-\frac{m_q}{N}\right)
\frac{N-V_R}{N-1}.
\tag{S3.5}
\end{equation}

Taking the square root of Eq. (S3.5) gives the exact relation

\begin{equation}
\frac{\sigma_T(\rho_R)}
     {\mathbb E_T[\rho_R]}
=
\lambda_R^{-1/2}
\left[
\left(1-\frac{m_q}{N}\right)
\frac{N-V_R}{N-1}
\right]^{1/2}
\le
\lambda_R^{-1/2}.
\tag{S3.6}
\end{equation}

In the sparse, nonsaturated regime \(m_q/N\to0\) and \(V_R/N\to0\), the
prefactor tends to one and the relative fluctuations scale as
\(\lambda_R^{-1/2}\). Away from full-graph saturation, large
\(\lambda_R\) therefore produces self-averaging occupancy, whereas for
\(\lambda_R=O(1)\) relative fluctuations remain of order one.

\hypertarget{si-3.2-why-the-accessibility-scale-is-a-mean-count-scale-not-a-sharp-transition}{%
\subsection{SI-3.2 Why the accessibility scale is a mean-count scale,
not a sharp
transition}\label{si-3.2-why-the-accessibility-scale-is-a-mean-count-scale-not-a-sharp-transition}}

The characteristic accessibility condition in the main text is

\begin{equation*}
\lambda_R=O(1).
\end{equation*}

This is not a thermodynamic phase transition. It is the scale at which
the expected number of targets in the accessible ball becomes order one.

A useful lower bound on the probability of at least one accessible
target follows from the standard second-moment argument. Since
\(m_R \geq 0\), the Cauchy--Schwarz inequality gives

\begin{equation*}
\Pr(m_R>0)
\ge
\frac{\mathbb E[m_R]^2}{\mathbb E[m_R^2]}.
\end{equation*}

Since

\begin{equation*}
\mathbb E[m_R^2]
=
\operatorname{Var}(m_R)+\lambda_R^2
\le
\lambda_R+\lambda_R^2,
\end{equation*}

we obtain

\begin{equation}
\Pr(m_R>0)
\ge
\frac{\lambda_R}{1+\lambda_R}.
\tag{S3.7}
\end{equation}

In particular, when the mean target count reaches one,

\begin{equation}
\lambda_R=1
\quad\Longrightarrow\quad
\Pr(m_R>0)\ge\frac12.
\tag{S3.8}
\end{equation}

This is why the criterion \(\lambda_R\sim1\) is a useful operational
notion of accessibility: at that point the probability of having at
least one target is already finite and bounded away from zero.

We can also control how many targets are present conditional on
accessibility. For any \(K>0\),

\begin{equation}
\Pr(m_R\ge K\mid m_R>0)
\le
\frac{\lambda_R}
     {K\Pr(m_R>0)}
\le
\frac{1+\lambda_R}{K}.
\tag{S3.9}
\end{equation}

Thus, when \(\lambda_R\) remains of order one, the conditional number of
targets also remains typically of order one.

In a more compact probabilistic notation this can be written as

\begin{equation}
m_R=O_p(1)
\qquad
\text{conditional on }m_R>0.
\tag{S3.10}
\end{equation}

No Poisson approximation is required for this statement. Under
additional sparse-limit assumptions one may obtain a Poisson limit, but
no main-text result depends on it.

\hypertarget{si-3.3-discrete-first-crossing-and-overshoot}{%
\subsection{SI-3.3 Discrete first crossing and
overshoot}\label{si-3.3-discrete-first-crossing-and-overshoot}}

On a finite graph we use

\begin{equation}
R_c^{\mathrm{op}}
=
\min\{R:\lambda_R\ge1\}.
\tag{S3.11}
\end{equation}

Because \(R\) is discrete, the mean target count need not be close to
one at the first crossing. On a rapidly expanding graph the ball can
jump from

\begin{equation*}
\lambda_{R_c^{\mathrm{op}}-1}<1
\end{equation*}

to a value well above one at the next shell.

Define the first-crossing overshoot factor

\begin{equation}
\omega_c
=
\lambda_{R_c^{\mathrm{op}}}
=
\frac{m_qV(R_c^{\mathrm{op}})}{N}
\ge1.
\tag{S3.12}
\end{equation}

The key question for the main text is whether this overshoot invalidates
the statement that the local target density at accessibility remains on
the scale \(m_q/N\). It does not.

For every radius,

\begin{equation*}
\mathbb E_T[\rho_R]
=
\frac{m_q}{N}.
\end{equation*}

Whenever \(\lambda_R\ge1\), Eq. (S3.7) gives \(\Pr(m_R>0)\ge1/2\).
Markov's inequality then yields, for every \(K>0\),

\begin{equation}
\Pr\left(
\rho_R>K\frac{m_q}{N}
\,\middle|\,
m_R>0
\right)
\le
\frac{2}{K}.
\tag{S3.13}
\end{equation}

This controls the upper tail of the local target density. A matching
lower-scale statement is also available.

Write

\begin{equation}
\frac{\rho_R}{m_q/N}
=
\frac{m_R}{\lambda_R}.
\tag{S3.14}
\end{equation}

If the overshoot \(\lambda_R\) stays bounded, then conditional on
accessibility \(m_R\ge1\), so the ratio in Eq. (S3.14) is bounded away
from zero by an order-one factor. If instead \(\lambda_R\) becomes
large, Eq. (S3.4) and Chebyshev's inequality give

\begin{equation}
\Pr\left(
\left|\frac{m_R}{\lambda_R}-1\right|>\varepsilon
\right)
\le
\frac{1}{\varepsilon^2\lambda_R},
\tag{S3.15}
\end{equation}

so \(m_R/\lambda_R\to1\) in probability. Since \(\Pr(m_R>0)\to1\) when
\(\lambda_R\to\infty\), conditioning on accessibility does not alter
this convergence. Combining the bounded- and large-overshoot cases shows
that, at the discrete first crossing,

\begin{equation}
\rho_{R_c^{\mathrm{op}}}
=
\Theta_p\!\left(\frac{m_q}{N}\right)
\qquad
\text{conditional on }m_{R_c^{\mathrm{op}}}>0.
\tag{S3.16}
\end{equation}

This is the precise sense in which the same target-density scale
survives discrete shell overshoot.

For the enrichment lower bound we only need the upper-tail control in
Eq. (S3.13). The exact enrichment requirement derived in SI-5 is

\begin{equation*}
E_R^{\min}
=
\frac{\delta}{M\rho_R}.
\end{equation*}

Combining this with Eq. (S3.13), with conditional probability at least
\(1-2/K\) one has

\begin{equation}
E_R^{\min}
\ge
\frac{\delta}{K}
\frac{N}{Mm_q}
=
\frac{\delta}{K\eta_q}.
\tag{S3.17}
\end{equation}

Because \(K\) can be chosen as any fixed constant, the minimum required
relevance enrichment is at least of order \(1/\eta_q\) with arbitrarily
high conditional probability as \(\eta_q\to0\). This is the precise
lower-bound statement behind the wording used in the main text.

Two limiting cases are worth distinguishing.

If the overshoot remains bounded, \(\omega_c=O(1)\), then the number of
targets at the first crossing remains typically order one.

If instead \(\omega_c\to\infty\), many targets are present and the local
density becomes self-averaging. In this case
\(\Pr(m_{R_c^{\mathrm{op}}}>0)\to1\), and conditional on accessibility,

\begin{equation}
\rho_{R_c^{\mathrm{op}}}
=
\frac{m_q}{N}
\left[
1+O_p(\omega_c^{-1/2})
\right].
\tag{S3.18}
\end{equation}

Therefore

\begin{equation}
E_{R_c^{\mathrm{op}}}^{\min}
=
\frac{\delta}{\eta_q}
\left[
1+O_p(\omega_c^{-1/2})
\right].
\tag{S3.19}
\end{equation}

Both bounded and large overshoot lead to the same leading enrichment
scale.

\hypertarget{si-3.4-a-secondary-count-scale-r_m}{%
\subsection{\texorpdfstring{SI-3.4 A secondary count scale
\(R_M\)}{SI-3.4 A secondary count scale R\_M}}\label{si-3.4-a-secondary-count-scale-r_m}}

It is sometimes useful to ask at what radius the accessible ball
contains not one target on average, but \(M\) targets on average. Define

\begin{equation*}
R_M:
\qquad
\bar m_{R_M}\sim M.
\end{equation*}

Equivalently,

\begin{equation}
V(R_M)
\sim
\frac{MN}{m_q}.
\tag{S3.20}
\end{equation}

This is \textbf{not} the accessibility radius and it is \textbf{not}
another searchability transition. Its meaning is statistical. Since
relative occupancy fluctuations scale as \(\lambda_R^{-1/2}\),

\begin{equation}
\frac{\sigma_T(\rho_{R_M})}
     {\mathbb E_T[\rho_{R_M}]}
=
O(M^{-1/2}).
\tag{S3.21}
\end{equation}

Thus \(R_M\) is simply a radius at which the accessible target count has
become sufficiently large that relative fluctuations are of order
\(M^{-1/2}\).

If \(M>m_q\), the condition \(\bar m_R=M\) cannot be reached even after
the whole graph is accessible, so \(R_M\) does not exist. None of the
results in the main text depends on this scale.

\hypertarget{si-3.5-fluctuations-of-the-minimum-enrichment}{%
\subsection{SI-3.5 Fluctuations of the minimum
enrichment}\label{si-3.5-fluctuations-of-the-minimum-enrichment}}

For every nonempty target realization,

\begin{equation}
E_R^{\min}
=
\frac{\delta \, V_R}{Mm_R}.
\tag{S3.22}
\end{equation}

Using

\begin{equation*}
\lambda_R=\frac{m_qV_R}{N},
\qquad
\eta_q=\frac{Mm_q}{N},
\end{equation*}

this can be rewritten as

\begin{equation}
E_R^{\min}
=
\frac{\delta}{\eta_q}
\frac{\lambda_R}{m_R}.
\tag{S3.23}
\end{equation}

When \(\lambda_R\gg1\), occupancy self-averages:

\begin{equation*}
m_R
=
\lambda_R
\left[
1+O_p(\lambda_R^{-1/2})
\right].
\end{equation*}

Hence

\begin{equation}
E_R^{\min}
=
\frac{\delta}{\eta_q}
\left[
1+O_p(\lambda_R^{-1/2})
\right].
\tag{S3.24}
\end{equation}

At \(R_M\), where \(\lambda_R\sim M\),

\begin{equation}
E_{R_M}^{\min}
=
\frac{\delta}{\eta_q}
\left[
1+O_p(M^{-1/2})
\right].
\tag{S3.25}
\end{equation}

The \(M^{-1/2}\) factor is therefore a fluctuation width around the
leading relevance-enrichment requirement. It is not a separate
transition.

\hypertarget{si-3.6-graph-to-graph-and-seed-to-seed-heterogeneity}{%
\subsection{SI-3.6 Graph-to-graph and seed-to-seed
heterogeneity}\label{si-3.6-graph-to-graph-and-seed-to-seed-heterogeneity}}

All preceding results in SI-3 conditioned on a fixed graph and seed. In
the numerical experiments, \(V_R\) itself varies across HRG realizations
and, within a heterogeneous graph, across seed vertices.

Conditional on a measured \(V_R\),

\begin{equation*}
\mathbb E_T[m_R\mid V_R]
=
\frac{m_q}{N}V_R.
\end{equation*}

The law of total variance separates target-placement fluctuations from
geometric fluctuations:

\begin{equation}
\operatorname{Var}_{G,T}(m_R)
=
\mathbb E_G[
\operatorname{Var}_T(m_R\mid V_R)
]
+
\operatorname{Var}_G[
\mathbb E_T(m_R\mid V_R)
].
\tag{S3.26}
\end{equation}

Substituting the hypergeometric variance gives

\begin{equation}
\operatorname{Var}_{G,T}(m_R)
=
\mathbb E_G\left[
V_R\frac{m_q}{N}
\left(1-\frac{m_q}{N}\right)
\frac{N-V_R}{N-1}
\right]
+
\left(\frac{m_q}{N}\right)^2
\operatorname{Var}_G(V_R).
\tag{S3.27}
\end{equation}

The first term is target-placement variability at fixed geometry. The
second is variability of the geometry itself.

On a periodic lattice, different seeds are equivalent and the geometric
term vanishes. In an HRG, ball growth depends strongly on the graph
realization and the seed.

For this reason, the operational accessibility radius is computed
separately for every graph and seed,

\begin{equation}
R_c^{\mathrm{op}}(G,x_0)
=
\min\left\{
R:
V_{G,x_0}(R)\ge\frac{N}{m_q}
\right\},
\tag{S3.28}
\end{equation}

and only then summarized across the ensemble.

This ordering matters. In general,

\begin{equation*}
\mathbb E_G[R_c^{\mathrm{op}}]
\end{equation*}

is not equal to the radius obtained by solving

\begin{equation*}
\mathbb E_G[V(R)]
=
\frac{N}{m_q},
\end{equation*}

because averaging and inversion need not commute.

\hypertarget{si-4-exact-finite-budget-neutral-benchmark}{%
\section{SI-4: Exact finite-budget neutral
benchmark}\label{si-4-exact-finite-budget-neutral-benchmark}}

\hypertarget{si-4.1-maximum-distinct-coverage}{%
\subsection{SI-4.1 Maximum distinct
coverage}\label{si-4.1-maximum-distinct-coverage}}

Suppose the accessible ball contains \(V_R\) vertices and the inspection
budget is \(M\). A target-blind search cannot inspect more than

\begin{equation}
D(R)=\min\{M,V_R\}
\tag{S4.1}
\end{equation}

distinct accessible vertices.

Under neutral target placement, all choices of \(D(R)\) distinct
vertices have the same target-averaged success probability. Therefore
the ideal strategy that uses every inspection to visit a new accessible
vertex has success probability

\begin{equation}
\Phi_0(R)
=
1-
\frac{\binom{N-D(R)}{m_q}}
     {\binom{N}{m_q}}.
\tag{S4.2}
\end{equation}

Equivalently,

\begin{equation}
\Phi_0(R)
=
1-
\frac{\binom{N-m_q}{D(R)}}
     {\binom{N}{D(R)}}.
\tag{S4.3}
\end{equation}

These expressions are exact at finite \(N\).

Equation (S4.2) should be read as an \textbf{ideal distinct-inspection
benchmark}. It is attained whenever the search mechanism is free to
choose \(D(R)\) distinct accessible vertices. A trajectory constrained
by local motion, revisits, or other dynamical restrictions may inspect
fewer distinct vertices and therefore performs no better than this
benchmark. This is why Eq. (S4.2) is the appropriate neutral upper
envelope for Fig. 1.

\hypertarget{si-4.2-two-regimes-accessibility-limited-and-budget-limited}{%
\subsection{SI-4.2 Two regimes: accessibility-limited and
budget-limited}\label{si-4.2-two-regimes-accessibility-limited-and-budget-limited}}

If

\begin{equation*}
V_R<M,
\end{equation*}

then the budget is large enough to inspect every accessible vertex. Thus

\begin{equation*}
D(R)=V_R
\end{equation*}

and

\begin{equation}
\Phi_0(R)
=
1-
\frac{\binom{N-V_R}{m_q}}
     {\binom{N}{m_q}}
\equiv
\Psi(R).
\tag{S4.4}
\end{equation}

In this regime, search fails only if the accessible ball contains no
target.

If instead

\begin{equation*}
V_R\ge M,
\end{equation*}

then only \(M\) of the accessible vertices can be inspected, so

\begin{equation*}
D(R)=M
\end{equation*}

and

\begin{equation}
\Phi_0(R)
=
1-
\frac{\binom{N-M}{m_q}}
     {\binom{N}{m_q}}.
\tag{S4.5}
\end{equation}

This expression no longer depends on \(R\). Once the accessible ball has
more than \(M\) vertices, increasing accessibility alone cannot improve
the neutral target-blind benchmark.

The crossover radius \(R_B\) is defined by

\begin{equation}
V(R_B)\sim M.
\tag{S4.6}
\end{equation}

If

\begin{equation*}
\eta_q=\frac{Mm_q}{N}\ll1,
\end{equation*}

then

\begin{equation*}
M\ll\frac{N}{m_q}.
\end{equation*}

Thus the budget threshold corresponds to a parametrically smaller
accessible volume than the characteristic accessibility threshold. On a
discrete graph the two thresholds can nevertheless be crossed within the
same shell if ball growth overshoots strongly. The key point is the
separation in volume: by the accessibility scale, a target-blind search
is already constrained by its finite inspection budget.

\hypertarget{si-4.3-sparse-large-n-form}{%
\subsection{\texorpdfstring{SI-4.3 Sparse large-\(N\)
form}{SI-4.3 Sparse large-N form}}\label{si-4.3-sparse-large-n-form}}

For \(D=D(R)\),

\begin{equation}
1-\Phi_0(R)
=
\prod_{j=0}^{D-1}
\left(
1-\frac{m_q}{N-j}
\right).
\tag{S4.7}
\end{equation}

If \(D=o(N)\), \(m_q=o(N)\), and \(Dm_q/N=O(1)\), then

\begin{equation}
\Phi_0(R)
\simeq
1-
\exp\left[
-\frac{D(R)m_q}{N}
\right].
\tag{S4.8}
\end{equation}

In the budget-limited regime, \(D(R)=M\), giving

\begin{equation}
\Phi_0
\simeq
1-e^{-\eta_q}.
\tag{S4.9}
\end{equation}

If \(\eta_q\ll1\),

\begin{equation}
\Phi_0
=
\eta_q
+
O(\eta_q^2).
\tag{S4.10}
\end{equation}

This exponential expression is simply the sparse asymptotic form of the
exact finite-population result. It is not based on an
independent-sampling assumption.

\hypertarget{si-4.4-sampling-with-replacement-as-a-comparison}{%
\subsection{SI-4.4 Sampling with replacement as a
comparison}\label{si-4.4-sampling-with-replacement-as-a-comparison}}

For intuition, suppose that each of the \(M\) inspections samples
uniformly \textbf{with replacement} from the accessible ball.
Conditional on \(m_R=x\) targets inside the ball,

\begin{equation}
\Phi_{\mathrm{wr}}(R\mid x)
=
1-
\left(
1-\frac{x}{V_R}
\right)^M.
\tag{S4.11}
\end{equation}

Under the neutral fixed-size target ensemble,

\begin{equation*}
m_R
\sim
\operatorname{Hypergeom}(N,m_q,V_R),
\end{equation*}

so averaging over target placement gives

\begin{equation}
\overline{\Phi}_{\mathrm{wr}}(R)
=
\sum_{x=x_{\min}}^{x_{\max}}
\left[
1-
\left(
1-\frac{x}{V_R}
\right)^M
\right]
\frac{
\binom{m_q}{x}\binom{N-m_q}{V_R-x}
}{
\binom{N}{V_R}
},
\tag{S4.12}
\end{equation}

where

\begin{equation*}
x_{\min}
=
\max\{0,V_R+m_q-N\},
\qquad
x_{\max}
=
\min\{V_R,m_q\}.
\end{equation*}

A with-replacement sample contains at most \(D(R)=\min\{M,V_R\}\)
distinct vertices. Conditional on its number \(d\) of distinct draws,
its target-averaged success probability is \(F(d)\), which is increasing
in \(d\). Therefore

\begin{equation}
\overline{\Phi}_{\mathrm{wr}}(R)
\le
\Phi_0(R).
\tag{S4.13}
\end{equation}

When \(M^2/V_R\to0\), repeated draws become rare and the
with-replacement and distinct-inspection benchmarks approach one
another.

Equation (S4.12) is included only as a comparison. Figure 1 uses the
exact distinct-inspection benchmark \(\Phi_0\) in Eq. (S4.2), not the
with-replacement quantity.

\hypertarget{si-4.5-no-additional-finite-m-transition}{%
\subsection{\texorpdfstring{SI-4.5 No additional finite-\(M\)
transition}{SI-4.5 No additional finite-M transition}}\label{si-4.5-no-additional-finite-m-transition}}

At finite \(N\), Eq. (S4.2) is an exact finite combinatorial expression;
the finite inspection budget introduces a coverage ceiling, not an
additional critical transition.

Two distinct finite-size effects should be kept separate:

\begin{enumerate}
\def\labelenumi{\arabic{enumi}.}
\tightlist
\item
  the inspection budget imposes the deterministic coverage ceiling
  \(D(R)=\min\{M,V_R\}\);
\item
  the number of targets inside the accessible ball fluctuates across
  target realizations, as analyzed in SI-3.
\end{enumerate}

The scale \(R_M\) uses \(M\) only as a reference occupancy level: the
associated \(M^{-1/2}\) width is a fluctuation scale, not a second
searchability transition.

\hypertarget{si-5-relevance-enrichment-recall-and-candidate-subgraphs}{%
\section{SI-5: Relevance enrichment, recall, and candidate
subgraphs}\label{si-5-relevance-enrichment-recall-and-candidate-subgraphs}}

This section gives the detailed version of the positive statement in the
main text:

\begin{quote}
if target-blind search can inspect only a vanishing fraction of the
characteristic accessible volume, finite-success search must
preferentially inspect vertices that are more likely to be relevant.
\end{quote}

The word \textbf{bias} is used here in a purely statistical sense:
inspection probabilities must be nonuniform in a way correlated with
relevance. It does not imply that any particular algorithmic bias is
desirable.

\hypertarget{si-5.1-average-probability-of-inspecting-a-target}{%
\subsection{SI-5.1 Average probability of inspecting a
target}\label{si-5.1-average-probability-of-inspecting-a-target}}

Fix a graph, seed, radius, and a nonempty target realization. Suppose
inspections remain inside \(B(x_0,R)\). Throughout SI-5, unless stated
otherwise, the graph, seed, radius, and target realization are fixed,
and probabilities refer to the randomness of the inspection mechanism.

For each inspection slot \(t=1,\ldots,M\), let

\begin{equation*}
H_t
=
\{\text{the vertex inspected at step }t\text{ is a target}\},
\end{equation*}

and denote

\begin{equation*}
p_{R,t}
=
\Pr(H_t).
\end{equation*}

If a practical search stops as soon as it finds a target, the later
slots can simply be ignored: the success event is still the union of the
hit events that actually occur. Equivalently, one may imagine continuing
the bookkeeping to \(M\) slots after a hit; this cannot change whether
at least one hit occurred. The argument below therefore does not rely on
independence between steps or on a particular stopping convention.

Define the average per-step target probability

\begin{equation}
p_R
=
\frac{1}{M}
\sum_{t=1}^M p_{R,t}.
\tag{S5.1}
\end{equation}

The probability of at least one hit is

\begin{equation*}
P_{\mathrm{succ}}
=
\Pr\left(\bigcup_{t=1}^M H_t\right).
\end{equation*}

By the union bound,

\begin{equation}
P_{\mathrm{succ}}
\le
\sum_{t=1}^M p_{R,t}
=
Mp_R.
\tag{S5.2}
\end{equation}

Thus finite success,

\begin{equation*}
P_{\mathrm{succ}}\ge\delta,
\end{equation*}

requires

\begin{equation}
p_R
\ge
\frac{\delta}{M}.
\tag{S5.3}
\end{equation}

Now compare this with uniform inspection of the full accessible ball.
For a fixed target realization, the target fraction in that ball is

\begin{equation*}
\rho_R
=
\frac{m_R}{V_R}.
\end{equation*}

A uniformly chosen accessible vertex is therefore a target with
probability \(\rho_R\). This motivates the \textbf{relevance-enrichment
factor}

\begin{equation}
E_R
=
\frac{p_R}{\rho_R}.
\tag{S5.4}
\end{equation}

For a full-budget policy that samples the accessible ball uniformly,
\(E_R=1\). Values \(E_R>1\) mean that the inspection mechanism gives
more exposure to relevant vertices than uniform inspection would.

Combining Eqs. (S5.3) and (S5.4) gives

\begin{equation}
E_R
\ge
\frac{\delta}{M\rho_R}
=
\frac{\delta V_R}{Mm_R}.
\tag{S5.5}
\end{equation}

This is the basic relevance-enrichment bound.

Its meaning can be read directly from the two factors. Finite success
requires an absolute per-step target probability of at least
\(\delta/M\). If the target fraction \(\rho_R\) in the accessible ball
is tiny, reaching that absolute probability requires a correspondingly
large enrichment relative to uniform inspection.

\hypertarget{si-5.2-expected-local-recall}{%
\subsection{SI-5.2 Expected local
recall}\label{si-5.2-expected-local-recall}}

Relevance enrichment is not the same thing as recall. To separate them,
let

\begin{equation*}
q_i
=
\Pr(i\text{ is inspected at least once})
\end{equation*}

for each accessible vertex.

The expected fraction of accessible targets that are reached at least
once is

\begin{equation}
A_R
=
\frac{1}{m_R}
\sum_{i\in T_q\cap B(x_0,R)}q_i.
\tag{S5.6}
\end{equation}

This is the expected \textbf{local recall} within the accessible ball.

Let \(K_R\) be the number of distinct target vertices inspected. Then

\begin{equation}
\mathbb E[K_R]
=
m_RA_R.
\tag{S5.7}
\end{equation}

Because

\begin{equation*}
P_{\mathrm{succ}}
=
\Pr(K_R\ge1)
\le
\mathbb E[K_R],
\end{equation*}

finite success requires

\begin{equation}
A_R
\ge
\frac{\delta}{m_R}.
\tag{S5.8}
\end{equation}

If only one or a few targets are accessible, a successful search must
give a non-negligible probability of reaching a substantial fraction of
them. If many targets are accessible, finite success can be obtained
with much smaller recall.

\hypertarget{si-5.3-candidate-sets-and-visible-subgraphs}{%
\subsection{SI-5.3 Candidate sets and visible
subgraphs}\label{si-5.3-candidate-sets-and-visible-subgraphs}}

Many practical search, ranking, and recommendation systems do not assign
a smoothly varying inspection probability to every accessible vertex.
Instead, they effectively restrict attention to a much smaller set of
candidates and then rank or inspect within that set.

To represent this situation, let

\begin{equation*}
S_R
\subseteq
B(x_0,R)
\end{equation*}

be a \textbf{candidate set}.

If one retains the graph edges whose endpoints both lie in \(S_R\), the
same vertex set defines an induced \textbf{candidate subgraph} or
\textbf{visible subgraph}. Connectivity is not required. The set
formulation is the more general one: a ranking or recommendation system
need not literally construct a graph at all. The mathematical point is
only that a much smaller collection of accessible alternatives receives
enough visibility to be considered for inspection.

Suppose, for the moment, that inspection is uniform within \(S_R\).
Define the fraction of the accessible ball that remains visible,

\begin{equation}
\phi_R
=
\frac{|S_R|}{V_R},
\tag{S5.9}
\end{equation}

the fraction of accessible targets retained in the candidate set,

\begin{equation}
a_R
=
\frac{|S_R\cap T_q|}{m_R},
\tag{S5.10}
\end{equation}

and the target fraction within the candidate set,

\begin{equation}
P_R
=
\frac{|S_R\cap T_q|}{|S_R|}.
\tag{S5.11}
\end{equation}

These quantities obey the identity

\begin{equation}
P_R
=
\frac{a_R}{\phi_R}\rho_R.
\tag{S5.12}
\end{equation}

The corresponding relevance enrichment is therefore

\begin{equation}
E_R
=
\frac{P_R}{\rho_R}
=
\frac{a_R}{\phi_R}.
\tag{S5.13}
\end{equation}

This equation makes the geometry transparent. To increase target density
relative to the full accessible ball, the algorithm must retain targets
more efficiently than it retains ordinary accessible vertices.

For a fixed candidate set \(S_R\), uniform inspection gives per-step
target probability \(P_R\). The union bound therefore gives

\begin{equation*}
P_{\mathrm{succ}}\le MP_R.
\end{equation*}

Hence finite success requires

\begin{equation}
P_R\ge\frac{\delta}{M}.
\tag{S5.14}
\end{equation}

This statement is conditional on the realized candidate set; if
candidate generation is itself random, the same bound applies to each
fixed realization of \(S_R\) for which success probability at least
\(\delta\) is required. Equation (S5.14) implies

\begin{equation}
\phi_R
\le
\frac{M}{\delta}a_R\rho_R.
\tag{S5.15}
\end{equation}

Because \(a_R\le1\),

\begin{equation*}
\phi_R
\le
\frac{M}{\delta}\rho_R.
\end{equation*}

Equivalently,

\begin{equation}
|S_R|
\le
\frac{M}{\delta}|S_R\cap T_q|
\le
\frac{Mm_R}{\delta}.
\tag{S5.16}
\end{equation}

At the operational first crossing, SI-3 gives

\begin{equation*}
\rho_{R_c^{\mathrm{op}}}
=
\Theta_p\!\left(\frac{m_q}{N}\right)
\end{equation*}

conditional on accessibility. Therefore any uniformly searched
candidate-set method achieving \(P_{\mathrm{succ}}\ge\delta\) satisfies

\begin{equation}
\phi_{R_c^{\mathrm{op}}}
=
O_p\!\left(\frac{\eta_q}{\delta}\right)
\qquad
\text{conditional on accessibility}.
\tag{S5.17}
\end{equation}

Equivalently, for every fixed \(K>0\), Eq. (S3.13) implies with
conditional probability at least \(1-2/K\) that

\begin{equation*}
\phi_{R_c^{\mathrm{op}}}
\le
\frac{K\eta_q}{\delta}.
\end{equation*}

Therefore, as

\begin{equation*}
\eta_q\to0,
\end{equation*}

any uniformly searched candidate set that maintains a fixed success
probability must occupy a \textbf{vanishing fraction of the accessible
region} in probability, conditional on accessibility.

This is the precise version of the visible-subgraph statement in the
main text:

\begin{equation*}
\boxed{
\text{accessible graph}
\;\longrightarrow\;
\text{relevance-selected candidate set}
\;\longrightarrow\;
\text{inspection}.
}
\end{equation*}

A real ranking or recommendation algorithm need not literally construct
a graph and then delete vertices. The mathematical abstraction is that
only a small subset of the accessible alternatives receives enough
visibility to be seriously considered or inspected.

If the first-crossing overshoot is bounded, then \(m_R=O_p(1)\)
conditional on accessibility, so Eq. (S5.16) also gives

\begin{equation*}
|S_R|=O_p(M).
\end{equation*}

This stronger statement about the \textbf{absolute} candidate-set size
is not needed in the main text because it depends on the overshoot
behavior. The vanishing-fraction statement is more robust.

\hypertarget{si-5.4-exact-success-within-a-candidate-set}{%
\subsection{SI-5.4 Exact success within a candidate
set}\label{si-5.4-exact-success-within-a-candidate-set}}

Let

\begin{equation*}
s=|S_R|,
\qquad
k=|S_R\cap T_q|.
\end{equation*}

Suppose

\begin{equation*}
d=\min\{M,s\}
\end{equation*}

distinct vertices are inspected uniformly from \(S_R\). The exact
success probability is

\begin{equation}
\Phi_S
=
1-
\frac{\binom{s-k}{d}}
     {\binom{s}{d}}.
\tag{S5.18}
\end{equation}

Since the candidate-set target fraction is \(P_R=k/s\), the union bound
gives

\begin{equation}
\Phi_S
\le
dP_R
\le
MP_R.
\tag{S5.19}
\end{equation}

Therefore

\begin{equation}
\Phi_S\ge\delta
\quad\Longrightarrow\quad
P_R\ge\frac{\delta}{M}.
\tag{S5.20}
\end{equation}

This conclusion is exact. It does not require sampling with replacement
and it does not require the sparse approximation.

If \(s\le M\), the entire candidate set can be inspected, so

\begin{equation}
\Phi_S
=
\mathbf 1_{\{k>0\}}.
\tag{S5.21}
\end{equation}

In that case the only remaining question is whether the
candidate-selection stage retained at least one target.

\hypertarget{si-5.5-relevance-enrichment-relative-to-uniform-visibility}{%
\subsection{SI-5.5 Relevance enrichment relative to uniform
visibility}\label{si-5.5-relevance-enrichment-relative-to-uniform-visibility}}

Uniform visibility over the full ball has target probability

\begin{equation*}
P_R^{(0)}=\rho_R.
\end{equation*}

Therefore

\begin{equation}
\frac{p_R}{P_R^{(0)}}
=
E_R.
\tag{S5.22}
\end{equation}

Every finite-success search obeys

\begin{equation}
E_R
\ge
\frac{\delta}{M\rho_R}.
\tag{S5.23}
\end{equation}

Define the minimum enrichment allowed by this necessary condition as

\begin{equation*}
E_R^{\min}
\equiv
\frac{\delta}{M\rho_R}.
\end{equation*}

At the operational first crossing, SI-3 gives

\begin{equation*}
\rho_{R_c^{\mathrm{op}}}
=
\Theta_p\!\left(\frac{m_q}{N}\right)
\end{equation*}

conditional on accessibility. Consequently,

\begin{equation}
E_{R_c^{\mathrm{op}}}^{\min}
=
\Theta_p\!\left(
\frac{\delta N}{Mm_q}
\right)
=
\Theta_p\!\left(
\frac{\delta}{\eta_q}
\right)
\qquad
\text{conditional on accessibility}.
\tag{S5.24}
\end{equation}

Every finite-success search must satisfy \(E_R\ge E_R^{\min}\). Thus
finite success requires relevance enrichment at least on the scale
\(1/\eta_q\); the actual enrichment may be larger.

The important divergence is therefore not necessarily in the
\textbf{absolute} target probability per inspection, which need only
exceed \(\delta/M\), but in the ratio between that probability and the
neutral baseline \(\rho_R\).

All enrichment and candidate-set bounds in this section are necessary
conditions for finite success, not sufficient ones. Satisfying them does
not by itself guarantee success, since correlations between inspections
or failures of candidate selection can further reduce the hit
probability.

\hypertarget{si-5.6-precisionrecall-geometry-of-candidate-selection}{%
\subsection{SI-5.6 Precision--recall geometry of candidate
selection}\label{si-5.6-precisionrecall-geometry-of-candidate-selection}}

For uniform inspection of a candidate set,

\begin{equation*}
P_R
=
\frac{a_R}{\phi_R}\rho_R.
\end{equation*}

Finite success requires

\begin{equation}
\frac{a_R}{\phi_R}
\ge
\frac{\delta}{M\rho_R}.
\tag{S5.25}
\end{equation}

At fixed enrichment \(E_R\),

\begin{equation}
P_R=E_R\rho_R,
\qquad
\phi_R=\frac{a_R}{E_R}.
\tag{S5.26}
\end{equation}

Thus increasing recall while keeping the same enrichment requires
increasing the visible fraction proportionally.

At fixed visible fraction,

\begin{equation}
P_R
=
\frac{\rho_R}{\phi_R}a_R.
\tag{S5.27}
\end{equation}

This makes explicit the tradeoff: when the neutral target density
\(\rho_R\) is extremely small, a candidate set cannot simultaneously
remain broad, provide only weak relevance enrichment, and still support
finite-budget success.

\hypertarget{si-5.7-retrieval-interpretation}{%
\subsection{SI-5.7 Retrieval
interpretation}\label{si-5.7-retrieval-interpretation}}

If \(M\) is interpreted as the number of results a user actually
inspects, Eq. (S5.3) says that success probability \(\delta\) requires
average target probability per inspected result at least

\begin{equation*}
\frac{\delta}{M}.
\end{equation*}

For example, if

\begin{equation*}
M=10,
\qquad
\delta=0.5,
\end{equation*}

the necessary average target probability is at least \(5\%\).

That number is not especially large by itself. The strong requirement
appears when the neutral baseline is tiny. If

\begin{equation*}
\rho_R\sim10^{-7},
\end{equation*}

then raising the target probability from \(10^{-7}\) to
\(5\times10^{-2}\) requires enrichment of at least

\begin{equation*}
5\times10^5.
\end{equation*}

This is why the main text emphasizes relevance enrichment rather than
absolute precision.

For retrieval in hyperbolic embedding spaces (Nickel and Kiela 2017;
Ganea, Bécigneul, and Hofmann 2018; Prokhorenkova et al. 2022), the same
counting logic applies after specifying a candidate region and a finite
inspection budget. The geometry that controls candidate accessibility
need not be the graph-distance geometry of the original network.

\begin{center}\rule{0.5\linewidth}{0.5pt}\end{center}

\hypertarget{summary-of-the-relevant-scales}{%
\section{Summary of the relevant
scales}\label{summary-of-the-relevant-scales}}

Three different scales appear in the analysis. Keeping them separate
prevents several possible confusions.

\hypertarget{accessibility-scale}{%
\subsection{Accessibility scale}\label{accessibility-scale}}

\begin{equation*}
V(R_c)
\sim
\frac{N}{m_q}.
\end{equation*}

This is the geometric scale at which the mean number of targets in the
accessible region becomes order one. On a discrete graph, the measured
first crossing is

\begin{equation*}
R_c^{\mathrm{op}}
=
\min\{R:\lambda_R\ge1\}.
\end{equation*}

\hypertarget{budget-saturation-scale}{%
\subsection{Budget-saturation scale}\label{budget-saturation-scale}}

\begin{equation*}
V(R_B)\sim M.
\end{equation*}

Beyond \(R_B\), a target-blind search cannot inspect the whole
accessible ball. Increasing \(R\) further does not improve the exact
neutral finite-budget benchmark \(\Phi_0\).

If \(\eta_q\ll1\), then

\begin{equation*}
M\ll\frac{N}{m_q},
\end{equation*}

so \(R_B\) is reached before \(R_c\).

\hypertarget{occupancy-self-averaging-scale}{%
\subsection{Occupancy self-averaging
scale}\label{occupancy-self-averaging-scale}}

When it exists,

\begin{equation*}
V(R_M)
\sim
\frac{MN}{m_q}.
\end{equation*}

At this scale the mean number of accessible targets is of order \(M\),
so relative target-count fluctuations are of order \(M^{-1/2}\). This is
a fluctuation scale, not a searchability threshold.

The central logic of the paper is therefore

\begin{equation*}
\boxed{
V(R_c)\sim\frac{N}{m_q}
}
\end{equation*}

for accessibility,

\begin{equation*}
\boxed{
P_{\mathrm{succ}}
\le
\eta_q
=
\frac{Mm_q}{N}
\sim
\frac{M}{V(R_c)}
}
\end{equation*}

for target-blind search, and, at the operational first crossing,

\begin{equation*}
\boxed{
E_{R_c^{\mathrm{op}}}^{\min}
=
\Theta_p\!\left(
\frac{\delta}{\eta_q}
\right)
}
\end{equation*}

conditional on accessibility, for the minimum relevance enrichment
compatible with finite success.

For uniformly searched candidate-set methods, the same result can be
expressed geometrically as

\begin{equation*}
\boxed{
\phi_{R_c^{\mathrm{op}}}
=
O_p\!\left(
\frac{\eta_q}{\delta}
\right)
}
\end{equation*}

conditional on accessibility and finite success. Thus, as
\(\eta_q\to0\), the visible candidate set must occupy a vanishing
fraction of the accessible region in probability.

\begin{center}\rule{0.5\linewidth}{0.5pt}\end{center}

\hypertarget{si-6-numerical-implementation-for-fig.-1}{%
\section{SI-6: Numerical implementation for Fig.
1}\label{si-6-numerical-implementation-for-fig.-1}}

\hypertarget{si-6.1-graph-ensembles}{%
\subsection{SI-6.1 Graph ensembles}\label{si-6.1-graph-ensembles}}

Figure 1 compares Krioukov hyperbolic random graphs (HRGs) with
two-dimensional periodic square lattices.

The HRG generator uses target mean degree

\begin{equation*}
\langle k\rangle=8,
\end{equation*}

degree exponent

\begin{equation*}
\gamma=2.5,
\end{equation*}

and temperature

\begin{equation*}
T=0.5.
\end{equation*}

The generated graph sizes are

\begin{equation*}
N_0
\in
\{
2500,
4900,
10^4,
22500,
4\times10^4,
9\times10^4
\}.
\end{equation*}

For each \(N_0\), 10 independent HRG realizations are generated.

The theoretical framework assumes a connected graph. Each generated HRG
is therefore restricted to its giant connected component (GCC). The
actual GCC size, rather than the generated size \(N_0\), is used as the
effective \(N\) in all target probabilities and accessibility criteria.
Across the ensembles used for Fig. 1, the GCC contains approximately
\(97\%\) of the generated vertices.

For each HRG realization, 30 seed vertices are sampled uniformly from
the GCC.

To keep the notation unambiguous, \(N_0\) denotes the
requested/generated HRG size, whereas \(N\) denotes the actual size of
the GCC used in the calculations. Panel (a) of Fig. 1 contains a
horizontal generated-size guide \(N_0/m_q\). By contrast,
\(R_c^{\mathrm{op}}\), \(\Psi(R)\), \(\Phi_0(R)\), and \(\eta_q\) are
evaluated using the actual \(N\) of each realization.

The control graph is an \(L\times L\) periodic square lattice with

\begin{equation*}
N=L^2
\end{equation*}

using the same generated sizes. Because the periodic lattice is
translation invariant, all seed vertices have the same graph-distance
ball volumes, so one seed is sufficient.

Before wrap-around effects,

\begin{equation}
V(R)
=
1+2R(R+1).
\tag{S6.1}
\end{equation}

\hypertarget{si-6.2-measuring-ball-growth-and-the-first-accessibility-crossing}{%
\subsection{SI-6.2 Measuring ball growth and the first accessibility
crossing}\label{si-6.2-measuring-ball-growth-and-the-first-accessibility-crossing}}

For every graph and seed, an exact breadth-first search determines the
number of vertices at each graph distance and therefore the cumulative
ball volume

\begin{equation*}
V_{G,x_0}(R)
=
|B_G(x_0,R)|.
\end{equation*}

The operational accessibility radius is computed \textbf{separately for
each graph and seed}:

\begin{equation}
R_c^{\mathrm{op}}(G,x_0)
=
\min\left\{
R:
V_{G,x_0}(R)
\ge
\frac{N}{m_q}
\right\}.
\tag{S6.2}
\end{equation}

Here \(N\) is the actual GCC size of that graph realization, not the
generated size \(N_0\).

Only after this first crossing has been found for each realization are
the values averaged.

This distinction is important because

\begin{equation*}
\text{mean of first-crossing radii}
\end{equation*}

and

\begin{equation*}
\text{first crossing of the mean ball-volume curve}
\end{equation*}

need not be the same quantity.

For each requested size \(N_0\) in Fig. 1(b), \(R_c^{\mathrm{op}}\) is
first averaged over seeds within each HRG realization. The plotted
ordinate is then the mean of these graph-level means, while the abscissa
is the corresponding graph-ensemble mean of the actual GCC scale
\(N/m_q\). Error bars denote standard deviations across graph-level
means after averaging over seeds within each graph and are smaller than
the symbol size. This preserves the graph/seed hierarchy rather than
treating all graph--seed pairs as independent realizations.

For the lattice, translation invariance makes the seed-level averaging
trivial.

\hypertarget{si-6.3-exact-accessibility-and-neutral-search-curves}{%
\subsection{SI-6.3 Exact accessibility and neutral-search
curves}\label{si-6.3-exact-accessibility-and-neutral-search-curves}}

The numerical example uses

\begin{equation*}
m_q=10,
\qquad
M=10.
\end{equation*}

No Monte Carlo sampling of target sets is needed. Once a graph-distance
ball volume \(V(R)\) has been measured, neutral target placement can be
integrated out exactly.

The accessibility probability is

\begin{equation}
\Psi(R)
=
1-
\frac{\binom{N-V(R)}{m_q}}
     {\binom{N}{m_q}},
\tag{S6.3}
\end{equation}

and the exact neutral distinct-inspection benchmark is

\begin{equation}
\Phi_0(R)
=
1-
\frac{
\binom{N-\min\{M,V(R)\}}{m_q}
}{
\binom{N}{m_q}
}.
\tag{S6.4}
\end{equation}

For every graph and seed, Eqs. (S6.3) and (S6.4) are evaluated using
that realization's actual GCC size \(N\) and measured ball volume
\(V(R)\). The resulting probabilities are averaged only afterwards. Thus
the plotted ensemble curves are averages of exact finite-population
probabilities, not probabilities evaluated from ensemble-averaged ball
volumes.

Figure 1(c) shows these ensemble averages for the HRG ensemble with
generated size

\begin{equation*}
N_0=4\times10^4.
\end{equation*}

Because the HRG degree distribution is broad, some high-degree seeds
already have

\begin{equation*}
V(1)>M.
\end{equation*}

For those seeds, \(\Phi_0(R)\) is already limited by the inspection
budget at \(R=1\), while \(\Psi(R)\) still depends on the full one-hop
neighborhood. This explains why the two ensemble-averaged curves can
begin to separate before the characteristic accessibility radius is
reached.

The effective GCC size is used separately for every graph realization.
Consequently

\begin{equation*}
\eta_q
=
\frac{Mm_q}{N}
\end{equation*}

varies slightly from graph to graph. For the \(N_0=4\times10^4\)
ensemble, its mean is approximately

\begin{equation*}
2.57\times10^{-3}.
\end{equation*}

\hypertarget{si-6.4-reproducibility}{%
\subsection{SI-6.4 Reproducibility}\label{si-6.4-reproducibility}}

The production data underlying Fig. 1 were generated on Windows 11 using

\begin{itemize}
\tightlist
\item
  Python 3.13.15,
\item
  NetworKit 11.2.1,
\item
  NumPy 2.5.2,
\item
  pandas 3.0.5.
\end{itemize}

A fixed master random seed, 20260823, controls graph generation and seed
selection.

Raw output is stored at graph--seed--radius resolution. Because the
neutral target ensemble is evaluated analytically rather than by Monte
Carlo target sampling, all quantities in Fig. 1 can be recomputed from
the saved graph sizes and ball volumes without regenerating target sets.

The numerical workflow is therefore:

\begin{enumerate}
\def\labelenumi{\arabic{enumi}.}
\tightlist
\item
  generate a graph realization;
\item
  restrict the HRG to its giant connected component;
\item
  choose seed vertices;
\item
  compute exact graph-distance ball volumes by breadth-first search;
\item
  determine \(R_c^{\mathrm{op}}\) separately for every graph and seed;
\item
  evaluate \(\Psi(R)\) and \(\Phi_0(R)\) exactly from the measured ball
  volumes;
\item
  preserve the graph/seed hierarchy when forming ensemble summaries.
\end{enumerate}

\hypertarget{refs}{}
\begin{CSLReferences}{1}{0}
\leavevmode\vadjust pre{\hypertarget{ref-Abdullah2017TypicalDistances}{}}%
Abdullah, Mohammed Amin, Nikolaos Fountoulakis, and Michel Bode. 2017.
{``Typical Distances in a Geometric Model for Complex Networks.''}
\emph{Internet Mathematics} 1 (1).
\url{https://doi.org/10.24166/im.13.2017}.

\leavevmode\vadjust pre{\hypertarget{ref-Boguna2009Ultrasmall}{}}%
Boguñá, Marián, and Dmitri Krioukov. 2009. {``Navigating Ultrasmall
Worlds in Ultrashort Time.''} \emph{Physical Review Letters} 102 (5):
058701. \url{https://doi.org/10.1103/PhysRevLett.102.058701}.

\leavevmode\vadjust pre{\hypertarget{ref-Boguna2009Navigability}{}}%
Boguñá, Marián, Dmitri Krioukov, and Kimberly C. Claffy. 2009.
{``Navigability of Complex Networks.''} \emph{Nature Physics} 5 (1):
74--80. \url{https://doi.org/10.1038/nphys1130}.

\leavevmode\vadjust pre{\hypertarget{ref-Fabbri2022ExposureInequality}{}}%
Fabbri, Francesco, Maria Luisa Croci, Francesco Bonchi, and Carlos
Castillo. 2022. {``Exposure Inequality in People Recommender Systems:
The Long-Term Effects.''} In \emph{Proceedings of the International AAAI
Conference on Web and Social Media}, 16:194--204. 1.
\url{https://ojs.aaai.org/index.php/ICWSM/article/view/19325}.

\leavevmode\vadjust pre{\hypertarget{ref-Ganea2018HyperbolicNN}{}}%
Ganea, Octavian-Eugen, Gary Bécigneul, and Thomas Hofmann. 2018.
{``Hyperbolic Neural Networks.''} In \emph{Advances in Neural
Information Processing Systems 31}, 5350--60. Curran Associates, Inc.
\url{https://papers.nips.cc/paper/7780-hyperbolic-neural-networks}.

\leavevmode\vadjust pre{\hypertarget{ref-Huszar2022TwitterAmplification}{}}%
Huszár, Ferenc, Sofia Ira Ktena, Conor O'Brien, Luca Belli, Adrian
Schlaikjer, Nicholas Hardt, Mark Jansen, Mykola R. Khodak, Levon
Aslanyan, and Amin Tootoonchian. 2022. {``Algorithmic Amplification of
Politics on Twitter.''} \emph{Proceedings of the National Academy of
Sciences of the United States of America} 119 (1): e2025334119.
\url{https://doi.org/10.1073/pnas.2025334119}.

\leavevmode\vadjust pre{\hypertarget{ref-Joachims2007Clicks}{}}%
Joachims, Thorsten, Laura Granka, Bing Pan, Helene Hembrooke, Filip
Radlinski, and Geri Gay. 2007. {``Evaluating the Accuracy of Implicit
Feedback from Clicks and Query Reformulations in Web Search.''}
\emph{ACM Transactions on Information Systems} 25 (2): 7.
\url{https://doi.org/10.1145/1229179.1229181}.

\leavevmode\vadjust pre{\hypertarget{ref-Kleinberg2000SmallWorld}{}}%
Kleinberg, Jon M. 2000. {``The Small-World Phenomenon: An Algorithmic
Perspective.''} In \emph{Proceedings of the Thirty-Second Annual ACM
Symposium on Theory of Computing}, 163--70. Association for Computing
Machinery. \url{https://doi.org/10.1145/335305.335325}.

\leavevmode\vadjust pre{\hypertarget{ref-Krioukov2010Hyperbolic}{}}%
Krioukov, Dmitri, Fragkiskos Papadopoulos, Maksim Kitsak, Amin Vahdat,
and Marián Boguñá. 2010. {``Hyperbolic Geometry of Complex Networks.''}
\emph{Physical Review E} 82 (3): 036106.
\url{https://doi.org/10.1103/PhysRevE.82.036106}.

\leavevmode\vadjust pre{\hypertarget{ref-Lattanzi2011Milgram}{}}%
Lattanzi, Silvio, Alessandro Panconesi, and D. Sivakumar. 2011.
{``Milgram-Routing in Social Networks.''} In \emph{Proceedings of the
20th International Conference on World Wide Web}, 725--34. Association
for Computing Machinery. \url{https://doi.org/10.1145/1963405.1963507}.

\leavevmode\vadjust pre{\hypertarget{ref-Nickel2017Poincare}{}}%
Nickel, Maximilian, and Douwe Kiela. 2017. {``Poincar{é} Embeddings for
Learning Hierarchical Representations.''} In \emph{Advances in Neural
Information Processing Systems 30}, 6338--47. Curran Associates, Inc.
\url{https://proceedings.neurips.cc/paper/2017/hash/59dfa2df42d9e3d41f5b02bfc32229dd-Abstract.html}.

\leavevmode\vadjust pre{\hypertarget{ref-Prokhorenkova2022HyperbolicNNS}{}}%
Prokhorenkova, Liudmila, Dmitry Baranchuk, Nikolay Bogachev, Yury
Demidovich, and Alexander Kolpakov. 2022. {``Graph-Based Nearest
Neighbor Search in Hyperbolic Spaces.''} In \emph{International
Conference on Learning Representations}.
\url{https://openreview.net/forum?id=USIgIY6TNDe}.

\end{CSLReferences}

\end{document}